\documentclass[onecolumn]{IEEEtran}
\usepackage{bibentry}
\usepackage{xcolor,soul,framed} 

\colorlet{shadecolor}{yellow}
\usepackage[pdftex]{graphicx}
\graphicspath{{../pdf/}{../jpeg/}}
\DeclareGraphicsExtensions{.pdf,.jpeg,.png}

\usepackage[cmex10]{amsmath}
\usepackage{array}
\usepackage{mdwmath}
\usepackage{mdwtab}
\usepackage{eqparbox}
\usepackage{url}
\usepackage{multirow}
\usepackage{float}
\usepackage{subfig}
\usepackage{lineno}
\begin{document}
\bstctlcite{IEEEexample:BSTcontrol}
    \title{Towards Data-Efficient Learning: A Benchmark for COVID-19 CT Lung and Infection Segmentation}
  \author{Jun Ma, Yixin Wang, Xingle An, Cheng Ge, Ziqi Yu, Jianan Chen, Qiongjie Zhu, Guoqiang Dong, Jian He, Zhiqiang He, Tianjia Cao, Yuntao Zhu, Ziwei Nie, Xiaoping Yang

\thanks{Jun Ma (corresponding author) is with Department of Mathematics, Nanjing University of Science and Technology, Nanjing, 210094, P. R. China (junma@njust.edu.cn).}
\thanks{Yixin Wang is with Institute of Computing Technology, Chinese Academy of Sciences; University of Chinese Academy of Sciences, Beijing, 100190, P. R. China (wangyixin19g@ict.ac.cn).}
\thanks{Xingle An is with China Electronics Cloud Brain (Tianjin) Technology CO., LTD, P. R. China (anxingle820@gmail.com).}
\thanks{Cheng Ge is with Institute of Bioinformatics and Medical Engineering, Jiangsu University of Technology, Changzhou, 213001, P. R. China.}
\thanks{Ziqi Yu is with Institute of Science and Technology for Brain-inspired Intelligence, Fudan University, Shanghai, P. R. China.}
\thanks{Jianan Chen is with Department of Medical Biophysics, University of Toronto, Toronto, ON, CA (chenjn2010@gmail.com).}
\thanks{Qiongjie Zhu, Guoqiang Dong, and Jian He is with Department of Radiology, Nanjing Drum Tower Hospital, the Affiliated Hospital of Nanjing University Medical School, Nanjing, 210008, P. R. China}
\thanks{Zhiqiang He is with Lenovo Ltd., Beijing, 100094, P. R. China.}
\thanks{Tianjia Cao is with China Electronics Cloud Brain (Tianjin) Technology CO., LTD, P. R. China.}
\thanks{Yuntao Zhu, Ziwei, Nie, and Xiaoping Yang (corresponding author) is with Department of Mathematics, Nanjing University, Nanjing, 210093, P. R. China (xpyang@nju.edu.cn).}
  }


\maketitle

\begin{abstract}
\noindent {\bf Purpose:}
Accurate segmentation of lung and infection in COVID-19 CT scans plays an important role in the quantitative management of patients. Most of the existing studies are based on large and private annotated datasets that are impractical to obtain from a single institution, especially when radiologists are busy fighting the coronavirus disease. Furthermore, it is hard to compare current COVID-19 CT segmentation methods as they are developed on different datasets, trained in different settings, and evaluated with different metrics.\\
{\bf Methods:}
To promote the development of data-efficient deep learning methods, in this paper, we built three benchmarks for lung and infection segmentation based on 70 annotated COVID-19 cases, which contain current active research areas, e.g., few-shot learning, domain generalization, and knowledge transfer.
For a fair comparison among different segmentation methods, we also provide standard training, validation and testing splits, evaluation metrics and, the corresponding code.\\
{\bf Results:}
Based on the state-of-the-art network, we provide more than 40 pre-trained baseline models, which not only serve as out-of-the-box segmentation tools but also save computational time for researchers who are interested in COVID-19 lung and infection segmentation. We achieve average Dice Similarity Coefficient (DSC) scores of 97.3\%, 97.7\%, and 67.3\% and average Normalized Surface Dice (NSD) scores of 90.6\%, 91.4\%, and 70.0\% for left lung, right lung, and infection, respectively. \\
{\bf Conclusions:}
To the best of our knowledge, this work presents the first data-efficient learning benchmark for medical image segmentation, and the largest number of pre-trained models up to now. All these resources are publicly available, and our work lays the foundation for promoting the development of deep learning methods for efficient COVID-19 CT segmentation with limited data.\\
\end{abstract}


%
\IEEEpeerreviewmaketitle


\section{Introduction}
\IEEEPARstart{C}{OVID-19} (coronavirus disease 2019) has spread all over the world during the past few months and caused over 61,000,000 people infected as of 30th November 2020 according to WHO statistics\footnote{https://covid19.who.int/}.
Computed tomography (CT) is playing an important role in the fight against COVID-19 \cite{RSNA_Expert_Consensus, RSNA_Expert_Consensus2, RSNA_Expert_Consensus3}. CT is shown to be more sensitive in the early diagnosis of COIVD-19 infection compared to Reverse Transcription-Polymerase Chain Reaction (RT-PCR) tests \cite{fang2020sensitivity}. Wang et al. trained a deep learning model on 325 COVID-19 CT scans and 740 typical pneumonia scans. Their model can identify 46 COVID-19 cases that were previously missed by the RT-PCR test \cite{wang2020deep}.  Further, quantitative information from CT images, such as the lung burden, the percentage of high opacity, and the lung severity score, can be used to monitor the disease progression and help us understand the course of COVID-19 \cite{COVID-19-Quant, chaganti2020quantification}.

Artificial Intelligence (AI) methods, especially deep learning-based methods, have been widely applied in medical image analysis to combat COVID-19 \cite{shen-review}. For example, AI can be used for building a contactless imaging workflow to prevent transmission from patients to health care providers \cite{wang2020precise}. In addition, most screening and segmentation algorithms for COVID-19 are developed with deep learning models, and the automatic diagnosis and COVID-19 infection quantification systems usually rely on the segmentation results generated by deep neural networks\cite{zheng2020deep,cao2020AILongAss,huang2020QuanCOVID, cell2020QuanCOVID}.

Although several studies show that deep learning methods have potential for providing accurate and quantitative assessment of COVID-19 infection in CT images \cite{shen-review}, the solutions mainly rely on large private datasets. Due to the patient privacy and intellectual property issues, the datasets and solutions may not be publicly available. However, researchers may hope that the datasets, entire source code and trained models could be provided by authors \cite{RSNAletterAIDis}.


Existing studies demonstrate that the classical U-Net (or V-Net) can achieve promising segmentation performance if hundreds of well-labelled training cases are available \cite{huang2020QuanCOVID, shan20quant}. 83.1\% to 91.6\% of segmentation performance in Dice score coefficient was reported in various U-Net-based approaches on different private datasets. Shan et al. developed a neural network based on V-Net and the bottle-neck structure to segment and quantify infection regions \cite{shan20quant,he2016deep}. In their human-in-the-loop strategy, they achieved 85.1\%, 91.0\% and 91.6\% dice with 36, 114, and 249 labelled CT scans, respectively.
Huang et al. \cite{huang2020QuanCOVID} employed U-Net \cite{ronneberger20152DUNet} for lung and infection using an annotated dataset of 774 cases, and demonstrated that the trained model could be used for quantifying the disease burden and monitoring disease progressions or treatment responses.
In general, we have observed a trend that training deep learning models with more annotations will decrease the time needed for contouring the infection, and increase segmentation accuracy, which is consistent with the Shan et al.'s findings \cite{shan20quant}. However, annotations for 3D CT volume data are expensive to acquire because it not only relies on professional diagnosis knowledge of radiologists, but also takes much time and labor, especially in current situation. Thus, a critical question would be:

\emph{how can we automatically annotate COVID-19 CT scans with limited
training data?}

Towards this question, three basic but important problems remain unsolved:
\begin{itemize}
    \item There is no publicly well-labelled COVID-19 CT 3D dataset; Data collection is the first and essential step to develop deep learning methods for COVID-19 segmentation. Most of the existing studies rely on large private dataset with hundreds of annotated CT scans.
    \item There are no public benchmarks to evaluate different deep learning-based solutions, and different studies often use various evaluation metrics. Similarly, datasets for training, validating and testing are split diversely, which also makes it hard for readers to compare those methods. For example, although MedSeg (http://medicalsegmentation.com/covid19/) dataset with 100 annotated slices were used in \cite{MedSeg100fan2020InfNet}, \cite{MedSeg100AttentionUNet}, \cite{MedSeg100MiniSeg}, \cite{MedSeg100MultiTask}, and \cite{MedSeg1002020ResAtten} to develop COVID-19 segmentation methods, it was split in different ways and the developed methods were evaluated by different metrics.
    \item There are no publicly available trained baseline models for COVID-19. U-Net, a well-known segmentation architecture, is commonly used as a baseline network in many studies. However, due to different implementations, the reported performance varies in different studies \cite{MedSeg100AttentionUNet, MedSeg1002020ResAtten}, even though the experiments are conducted on the same dataset.
\end{itemize}

In this paper, we focus on annotation-efficient deep learning solutions and aim to alleviate the above problems by providing a well-labelled COVID-19 CT dataset and a benchmark. In particular, we first provide a COVID-19 3D CT dataset with left lung, right lung, and infection annotations, and then establish three benchmark tasks to explore different deep learning strategies with limited training cases. Finally, we build comprehensive baselines for each task based on U-Net.

Our tasks target on three popular research fields in medical imaging community:
\begin{itemize}
    \item Few-shot learning: building high performance models from very few samples. Although most existing approaches focus on natural images, it has received growing attention in medical imaging because generating labels for medical images is much more difficult \cite{Zhao_2019_CVPR,roy2020MIAsqueeze}.
    \item Domain generalization: learning from known domains and applying to an unknown target domain that has different data distribution \cite{li2019episodic}. The ultimate goal of domain generalization is to train robust models that are generalizable to new unseen domains. Recently, model-agnostic learning has been an emerging research topic to achieve this goal \cite{DomainGen19DouQi}.
    \item Knowledge transfer: reusing existing annotations to boost the training/fine-tuning on a new dataset or a related new task. In contrast to domain generalization, both the source domain/task and the target domain/task are known, and we focus on storing and reusing knowledge gained from the source domain/task.
    This task has achieved significant attention in recent studies, such as using transfer learning or generative adversarial network to transfer knowledge from publicly available annotated datasets to new datasets \cite{PKUMICCAI19CardiacVessel, miccai_simu2019cell} in segmentation tasks.
\end{itemize}

Large-scale data remedies of lung and infection segmentation have been well studied \cite{huang2020QuanCOVID,cell2020QuanCOVID}. In this paper, we focus on small-data learning tasks because it is a more practical problem and large-scale annotated datasets are expensive and time-consuming to collect.
Moreover, the goal is to lay the foundation for important machine learning tasks when only limited cases are available. Designing novel methods is beyond the scope of this paper.
Our contributions can be summarized as follows.
\begin{itemize}
    \item We present a new COVID-19 CT datasets and the left lung, the right lung, and the infections are well annotated by chest radiologists.
    \item We set up 3 benchmark tasks to promote the studies on data-efficient deep learning for COVID-19 CT scans segmentation. Specifically, we focus on few-shot learning, domain generalization, and knowledge transfer that are also current research hotspots. To the best of our knowledge, this is the first data-efficient learning benchmark in medical image segmentation.
    \item We provide 40+ trained state-of-the-art models and corresponding segmentation results are publicly available, which can serve as strong baselines. More importantly, these trained models can be used as out-of-the-box tools for COVID-19 CT lung and infection segmentation, which could reduce the annotation time for radiologists.
\end{itemize}

\section{Materials}
\begin{figure*}[htbp]
\centering
\includegraphics[scale=0.6]{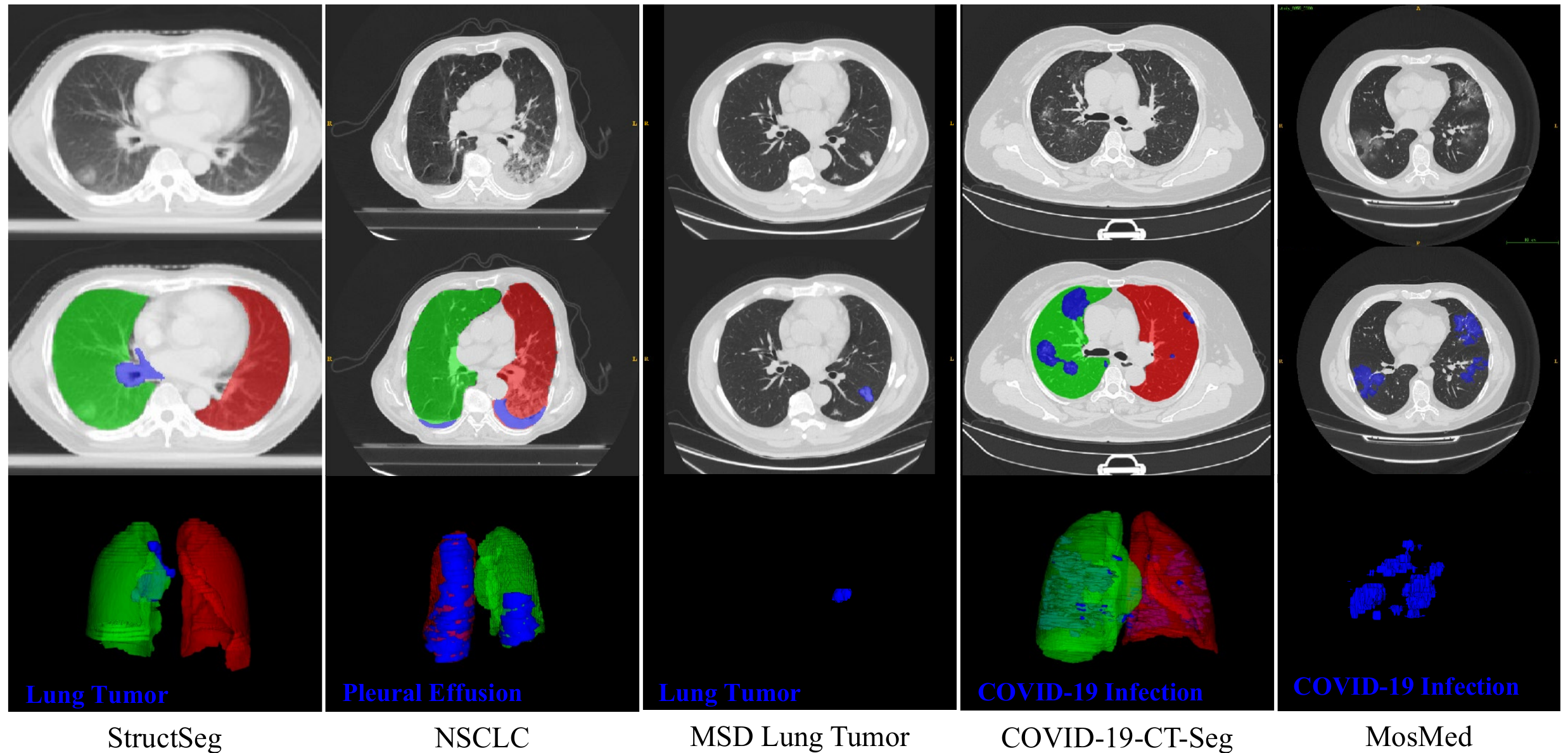}
\caption{Examples of five lung CT datasets. The 1st and 2nd row denote original non-contrast CT images and corresponding ground truth of lung and lesions, respectively. The 3rd row shows the 3D rendering results of ground truth. MSD Lung Tumor and MosMed datasets (3rd column) do not provide lung masks. The red, green, blue color denote left lung, right lung, and lung lesions, respectively. The blue legend in the 3rd rows stands for different lung lesion types.}
\label{fig:dataexample}
\end{figure*}

Annotations of COVID-19 CT scans are scarce, but several lung CT annotations with other diseases are publicly available. Thus, one of the main goals of our benchmark is to explore whether it is possible for using these existing annotations to assist COVID-19 CT segmentation.
This section introduces the public datasets used in our segmentation benchmarks. Figure \ref{fig:dataexample} presents some examples from each dataset.

\subsection{Existing lung CT segmentation datasets}

\subsubsection{StructSeg lung organ segmentation} 50 lung cancer patient CT scans are accessible, and all the cases are from one medical center. This dataset served as a segmentation challenge\footnote{MICCAI 2019 StructSeg: https://structseg2019.grand-challenge.org} during MICCAI 2019. Six organs are annotated, including left lung, right lung, spinal cord, esophagus, heart, and trachea. In this paper, we only use the left lung and right lung annotations.

\subsubsection{NSCLC left and right lung segmentation} This dataset consists of left and right thoracic volume segmentations delineated on 402 CT scans from The Cancer Imaging Archive NSCLC Radiomics \cite{NSCLC1, NSCLC2, TCIA}.

\subsection{Existing lung lesion CT segmentation datasets}
\subsubsection{MSD Lung tumor segmentation} This dataset is comprised of patients with non-small cell lung cancer from Stanford University (Palo Alto, CA, USA) publicly available through TCIA. The dataset served as a segmentation challenge\footnote{MICCAI 2018 MSD: http://medicaldecathlon.com/} during MICCAI 2018. The tumor is annotated by an expert thoracic radiologist, and 63 labelled CT scans are available.

\subsubsection{StructSeg Gross Target Volume segmentation of lung cancer} The same 50 lung cancer patient CT scans as the above StructSeg lung organ segmentation dataset are provided, and gross target volumes of tumors are annotated in each case.

\subsubsection{NSCLC Pleural Effusion (PE) segmentation} The CT scans in this dataset are the same as those in NSCLC left and right lung segmentation dataset, while pleural effusion is delineated for 78 cases \cite{NSCLC1, NSCLC2, TCIA}.

\subsubsection{MosMed Dataset} This dataset contains 50 annotated COVID-19 CT scans that are provided by municipal hospitals in Moscow, Russia \cite{MosMedData}. To evaluate the generalization ability of deep learning models, we used this dataset as an independent testing set in following benchmark settings.

\subsection{Our COVID-19-CT-Seg dataset}
We collected 20 public COVID-19 CT scans from the Coronacases Initiative and Radiopaedia, which can be freely downloaded\footnote{https://github.com/ieee8023/covid-chestxray-dataset} with CC BY-NC-SA license. All the cases contain COVID-19 infections. The proportion of infections in the lungs ranges from 0.01\%-59\%.
The left lung, right lung, and infection\footnote{The infection means to include all visibly affected regions of the lungs.} were firstly delineated by junior annotators with 1-5 years experience, then refined by two radiologists with 5-10 years experience, and finally all the annotations were verified and refined by a senior radiologist with more than 10 years experience in chest radiology. The whole lung mask includes both normal and pathological regions. All the annotations were manually performed by ITK-SNAP in a slice-by-slice manner on axial images. On average, it takes about $400 \pm 45$ minutes to delineate one CT scan with 250 slices.
There are totally $300+$ infections with $1800+$ slices.
We have made all the annotations publicly available \cite{COVID-19-CT-Seg} at \url{https://zenodo.org/record/3757476} with CC BY-NC-SA license.

\section{Methods}
As mentioned in Section 1, there is a need for innovative strategies that enable data-efficient methods for COVID-CT segmentation. Thus, we set up three tasks to evaluate potential annotation-efficient strategies.
In particular, we focus on learning to segment left lung, right lung and infection in COVID-19 CT scans using
\begin{itemize}
    \item pure but limited COVID-19 CT scans;
    \item existing annotated lung CT scans from other non-COVID-19 lung diseases;
    \item heterogeneous datasets include both COVID-19 and non-COVID-19 CT scans.
\end{itemize}
Furthermore, we also provide unified data (training, validation, and testing) splits, experimental settings, and evaluation metrics to standardize the deep learning-based segmentation protocols that can enable fair comparisons between different studies.

\subsection{Task 1: Learning with limited annotations}
Task 1 (Table \ref{tab:FSL}) is designed to address the problem of few-shot learning, where few annotations are available for training. This task is based on the COVID-19-CT-Seg dataset only. It contains three subtasks aiming to segment lung, infection and both of them, respectively. For each subtask, 5-fold cross validation results (based on a pre-defined dataset split file) are reported. In each fold, 4 training cases are used for training and the rest 16 cases are used for validation. Moreover, MosMed dataset is used as an independent testing set.

\begin{table}[htbp]
\centering
\caption{Experimental settings of Task 1 (Learning with limited annotations) for lung and infection segmentation in COVID-19 CT scans. All the experiments are base on the COVID-19-CT-Seg dataset. (Number) denotes the number of cases in the dataset.}
\label{tab:FSL}
\begin{tabular}{c|c|c}
\hline
Seg. Task          & Training and Validation                                                                                                                        & Testing                  \\
\hline
Lung               & \multirow{3}{*}{\begin{tabular}[c]{@{}c@{}}5-fold cross validation\\4 cases (20\% for training)\\16 cases (80\% for validation) \end{tabular}} & \multirow{3}{*}{MosMed (50)}  \\
\cline{1-1}
Infection          &                                                                                                                                                &                          \\
\cline{1-1}
Lung and infection &                                                                                                                                                &                          \\
\hline
\end{tabular}
\end{table}


\begin{table*}[!h]
\centering
\caption{Experimental settings of Task 2 (Learning to segment COVID-19 CT scans from non-COVID-19 CT scans) for lung and infection segmentation in COVID-19 CT scans. (Number) denotes the number of cases in the dataset.}
\label{tab:DG}
\begin{tabular}{c|l|l|c|c}
\hline
Seg. Task                  & \multicolumn{1}{c|}{Training} & \multicolumn{1}{c|}{In-domain Testing}                      & (Unseen) Testing 1 & (Unseen) Testing 2            \\
\hline
\multirow{2}{*}{Lung}      & StructSeg Lung (40)           & StructSeg Lung (10)                                         & \multirow{2}{*}{\begin{tabular}[c]{@{}c@{}}COVID-CT-Seg \\ Lung (20) \end{tabular}}          & \multirow{2}{*}{-}            \\
                           & NSCLC Lung (322)              & NSCLC Lung (80)                                             &                                                                                              &                               \\
\hline
\multirow{3}{*}{Infection} & MSD Lung Tumor (51)           & MSD Lung Tumor (12)                                         & \multirow{3}{*}{\begin{tabular}[c]{@{}c@{}}COVID-CT-Seg \\ Infection (20) \end{tabular}}     & \multirow{3}{*}{MosMed (50)}  \\
                           & StructSeg Gross Target (40)   & StructSeg Gross Target (10) &                                                                                              &                               \\
                           & NSCLC Pleural Effusion (62)   & NSCLC Pleural Effusion (16)                                 &                                                                                              &                               \\
\hline
\end{tabular}
\end{table*}

\begin{table*}[!htbp]
\caption{Experimental settings of Task 3 (Learning with both COVID-19 and non-COVID-19 CT scans) for lung and infection segmentation in COVID-19 CT scans. (Number) denotes the number of cases in the dataset.}\label{tab:TL}
\centering
\begin{tabular}{c|l|c|l|c|c}
\hline
Seg. Task                  & \multicolumn{2}{c|}{Training}                                                                                       & \multicolumn{1}{c|}{Validation}             & \multicolumn{1}{l|}{Testing 1}                                                           & Testing 2                     \\
\hline
\multirow{2}{*}{Lung}      & StructSeg Lung (40)         & \multirow{2}{*}{\begin{tabular}[c]{@{}c@{}}COVID-CT-Seg\\Lung (4) \end{tabular}}      & StructSeg Lung (10)                         & \multirow{2}{*}{\begin{tabular}[c]{@{}c@{}}COVID-CT-Seg \\ Lung (16) \end{tabular}}      & \multirow{2}{*}{-}            \\
                           & NSCLC Lung (322)            &                                                                                       & NSCLC Lung (80)                             &                                                                                          &                               \\
\hline
\multirow{3}{*}{Infection} & MSD Lung Tumor (51)         & \multirow{3}{*}{\begin{tabular}[c]{@{}c@{}}COVID-CT-Seg\\Infection (4) \end{tabular}} & MSD Lung Tumor (12)                         & \multirow{3}{*}{\begin{tabular}[c]{@{}c@{}}COVID-CT-Seg \\ Infection (16) \end{tabular}} & \multirow{3}{*}{MosMed (50)}  \\
                           & StructSeg Gross Target (40) &                                                                                       & StructSeg Gross Target (10) &                                                                                          &                               \\
                           & NSCLC Pleural Effusion (62) &                                                                                       & NSCLC Pleural Effusion (16)                 &                                                                                          &                               \\
\hline
\end{tabular}
\end{table*}

\subsection{Task 2: Learning to segment COVID-19 CT scans from non-COVID-19 CT scans}

Task 2 (Table \ref{tab:DG}) is designed to address the problem of domain generalization, where only out-of-domain data (non-COVID-19 datasets) are available for training. Specifically, in the first subtask, the StructSeg Lung dataset and the NSCLC Lung dataset are used for training. In the second subtask, the MSD Lung Tumor, the StructSeg Gross Target and the NSCLC Pleural Effusion datasets are used as training sets. For both subtasks, 80\% of the data are randomly selected for training and the rest 20\% are held-out as in-domain testing sets. All cases in two labelled COVID-19 CT datasets are kept for testing.

\subsection{Task 3: Learning with both COVID-19 and non-COVID-19 CT scans}
Task 3 (Table \ref{tab:TL}) is designed to address the problem of knowledge transfer with heterogeneous datasets, where both in-domain and out-of-domain data are included in the training set. Specifically, in both subtasks (lung segmentation and lung infection segmentation), 80\% non-COVID-19 data and 20\% COVID-19 data are used for training, while remained 20\% and 80\% data are used for validation and testing, respectively. Besides, MosMed dataset is used as an additional testing set.

\subsection{Evaluation metrics}
Motivated by the evaluation methods of the well-known medical image segmentation decathlon\footnote{http://medicaldecathlon.com/files/MSD-Ranking-scheme.pdf}, we also employ two complementary metrics to evaluate the segmentation performance. Dice similarity coefficient, a region-based measure, is used to evaluate the region overlap. Normalized surface Dice \cite{nikolov2018SDice}, a boundary-based measure is used to evaluate how close the segmentation and ground truth surfaces are to each other at a specified tolerance $\tau$.
For both two metrics, higher scores admit better segmentation performance, and $100\%$ means perfect segmentation.
Let $G, S$ denote the ground truth and the segmentation result, respectively. We formulate the definitions of the two measures as follows:
\subsubsection{Region-based measure}
\begin{equation*}
    DSC(G, S) = \frac{2|G\cap S|}{|G| + |S|};
\end{equation*}

\subsubsection{Boundary-based measure}
\begin{equation*}
    NSD(G, S) = \frac{|\partial G\cap B_{\partial S}^{(\tau)}| + |\partial S\cap B_{\partial G}^{(\tau)}|}{|\partial G| + |\partial S|}
\end{equation*}
where $B_{\partial G}^{(\tau)}, B_{\partial S}^{(\tau)} \subset R^3$  denote the border region of ground truth and segmentation surface at tolerance $\tau$, which are defined as $B_{\partial G}^{(\tau)} = \{x\in R^3 \, | \, \exists \tilde{x}\in \partial G,\, ||x-\tilde{x}||\leq \tau \}$ and $B_{\partial S}^{(\tau)} = \{x\in R^3 \,|\, \exists \tilde{x}\in \partial S,\, ||x-\tilde{x}||\leq \tau \}$, respectively. We set tolerance $\tau$ as $1mm$ and $3mm$ for lung segmentation and infection segmentation, respectively. The tolerance is computed by measuring the inter-rater segmentation variation between two different radiologists, which is also in accordance with another independent study \cite{nikolov2018SDice}. Python implementations of the two metrics are publicly available\footnote{https://github.com/JunMa11/COVID-19-CT-Seg-Benchmark/blob/master/utils/COVID-19-Seg-Evaluation.py}.

The main benefit of introducing Surface Dice is that it ignores small boundary deviations because small inter-observer errors are also unavoidable and often not clinically relevant when segmenting the objects by radiologists.

\subsection{U-Net baselines: oldies but goldies}
U-Net (\cite{ronneberger20152DUNet, ronneberger20163DUNet}) has been proposed for 5 years, and many variants have been proposed to improve it. However, recent study \cite{nnUNet2020} demonstrates that it is still hard to surpass a basic U-Net if the corresponding pipeline is designed adequately. In particular, nnU-Net (no-new-U-Net)  \cite{nnUNet2020} was proposed to automatically adapt preprocessing strategies and network architectures (i.e., the number of pooling, convolutional kernel size, and stride size) to a given 3D medical dataset. Without manual tuning, nnU-Net can achieve better performance than most specialised deep learning pipelines in 19 public international segmentation competitions and set a new state-of-the-art in the majority of 49 tasks. The source code is publicly available at \url{https://github.com/MIC-DKFZ/nnUNet}.

\begin{figure}[!htbp]
\begin{center}
   \includegraphics[scale=0.4]{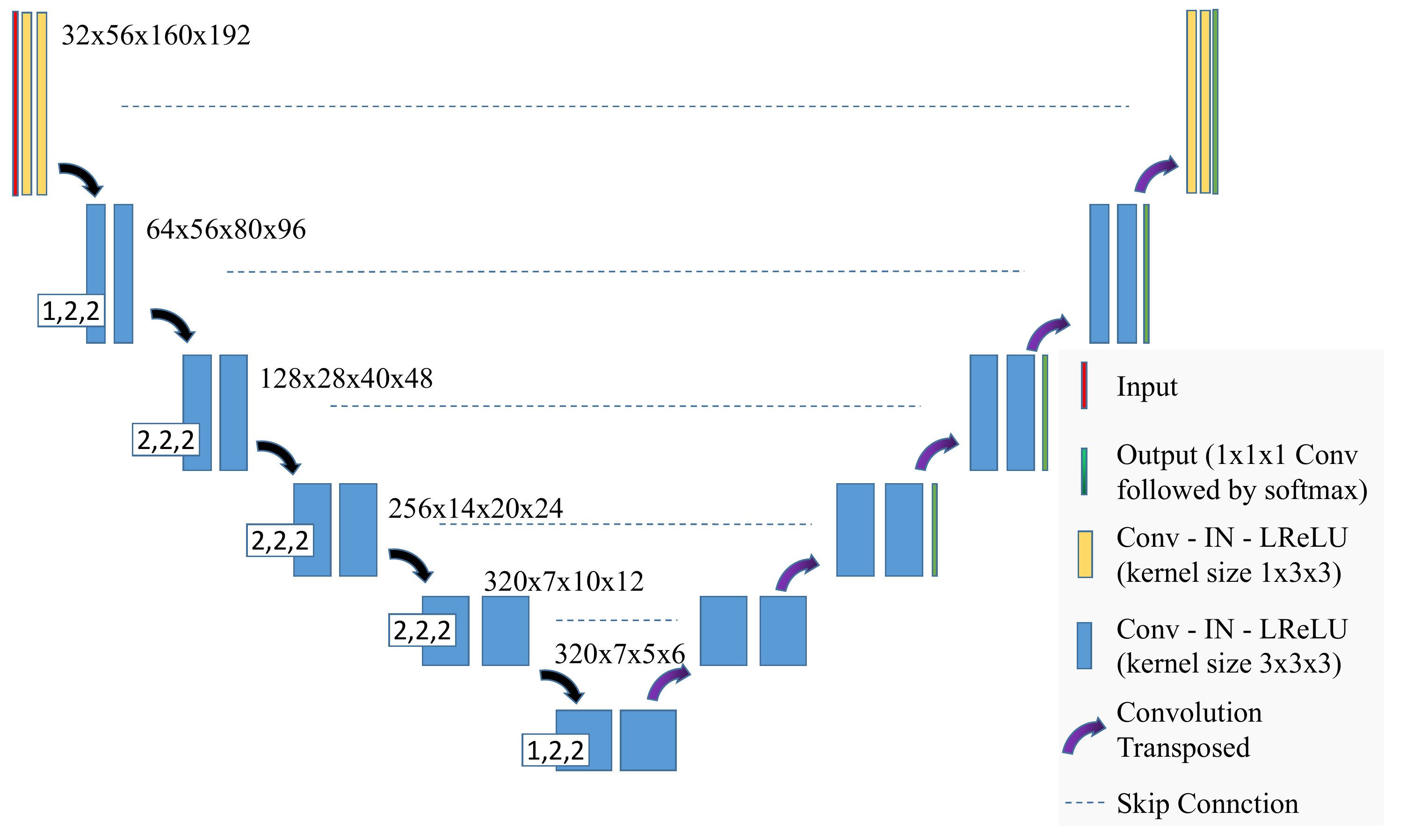}
    \caption{Details of the 3D U-Net architecture that is used in this work. The numbers (e.g. $32\times56\times160\times92$) near convolutional blocks denote feature map size in each resolution, and the numbers in white rectangles (e.g., $1,2,2$) denote stride sizes of convolutional kernels.}\label{fig:unet}
\end{center}
\end{figure}

\begin{table*}[!htbp]
\caption{Quantitative Results of 5-fold cross validation on COVID-19-CT-Seg dataset for Task 1: Learning with limited annotations. For each fold, average DSC and NSD values are reported. The last row shows the average results of 80 ($=5$ folds $\times 16$ testing cases per fold) testing cases.}\label{tab:exp-1}
\centering
\resizebox{\textwidth}{15mm}{
\begin{tabular}{c|cc|cc|cc|cc|cc|cc}
\hline
\multirow{3}{*}{Subtask} & \multicolumn{4}{c|}{Lung}                                        & \multicolumn{2}{c|}{\multirow{2}{*}{Infection}} & \multicolumn{6}{c}{Lung and Infection Union Segmentation}                                                            \\
\cline{2-5}\cline{8-13}
                         & \multicolumn{2}{c|}{Left Lung} & \multicolumn{2}{c|}{Right Lung} & \multicolumn{2}{c|}{}                           & \multicolumn{2}{c|}{Left Lung} & \multicolumn{2}{c|}{Right Lung} & \multicolumn{2}{c}{Infection}  \\
\cline{2-13}
                         & DSC (\%)                & NSD (\%)       & DSC (\%) & NSD (\%)                       & DSC (\%) & NSD (\%)                                       & DSC (\%) & NSD (\%)                      & DSC (\%) & NSD (\%)                       & DSC (\%) & NSD (\%)                      \\
\hline
Fold-0  & 84.9 $\pm$ \ 8.2  & 68.7 $\pm$ 13.3  & 85.2 $\pm$ 13.0  & 70.6 $\pm$ 15.8 & 68.1 $\pm$ 20.5 & 70.9 $\pm$ 21.3  & 50.5 $\pm$ 30.4 & 36.9 $\pm$ 19.6 & 64.8 $\pm$ 18.9 & 47.1 $\pm$ 13.8  & 66.5 $\pm$ 23.4 & 68.7 $\pm$ 22.5  \\
Fold-1  & 80.3 $\pm$ 14.5  & 61.8 $\pm$ 15.1  & 83.9 $\pm$ \ 9.6  & 68.3 $\pm$ \ 9.0 & 71.3 $\pm$ 20.5 & 71.8 $\pm$ 23.0  & 40.3 $\pm$ 18.7 & 27.5 $\pm$ 12.0 & 60.1 $\pm$ 11.1 & 41.7 $\pm$ \ 9.9  & 64.7 $\pm$ 21.8 & 60.6 $\pm$ 25.1  \\
Fold-2  & 87.1 $\pm$ 12.1  & 74.3 $\pm$ 16.0  & 90.3 $\pm$ \ 8.2  & 78.5 $\pm$ 12.0 & 66.2 $\pm$ 21.7 & 71.7 $\pm$ 24.2  & 80.3 $\pm$ 18.8 & 66.8 $\pm$ 18.8 & 85.2 $\pm$ 12.4 & 68.6 $\pm$ 15.1  & 60.7 $\pm$ 27.6 & 62.5 $\pm$ 28.9  \\
Fold-3  & 88.4 $\pm$ \ 7.0  & 75.2 $\pm$ \ 8.8  & 89.9 $\pm$ \ 6.3  & 78.5 $\pm$ \ 8.0 & 68.1 $\pm$ 23.1 & 70.8 $\pm$ 27.1  & 79.7 $\pm$ 13.6 & 65.4 $\pm$ 14.4 & 84.0 $\pm$ \ 9.8 & 67.7 $\pm$ 13.0  & 62.0 $\pm$ 27.9 & 65.3 $\pm$ 28.9  \\
Fold-4  & 88.3 $\pm$ \ 7.6  & 75.8 $\pm$ 11.0  & 90.2 $\pm$ \ 7.0  & 78.3 $\pm$ 10.2 & 62.7 $\pm$ 26.9 & 64.9 $\pm$ 28.2  & 72.4 $\pm$ 21.1 & 58.6 $\pm$ 20.8 & 80.9 $\pm$ 13.4 & 63.4 $\pm$ 15.9  & 51.4 $\pm$ 30.2 & 51.9 $\pm$ 31.0  \\ \hline
Avg   & \textbf{85.8 $\pm$ 10.5}  & \textbf{71.2 $\pm$ 13.8}  & \textbf{87.9 $\pm$ \ 9.3}   & \textbf{74.8 $\pm$ 11.9} & \textbf{67.3 $\pm$ 22.3} & \textbf{70.0 $\pm$ 24.4}  & 64.6 $\pm$ 26.4 & 51.1 $\pm$ 23.4 & 75.0 $\pm$ 16.8 & 57.7 $\pm$ 17.4  & 61.0 $\pm$ 26.2 & 61.8 $\pm$ 27.4     \\ \hline
\end{tabular}}
\end{table*}

\begin{table*}[!htbp]
\caption{Quantitative results (mean $\pm$ standard deviation) of Lung segmentation in Task 2.}\label{tab:t2-lung}
\centering
\begin{tabular}{c|cccc|cccc}
\hline
\multirow{3}{*}{Subtask} & \multicolumn{4}{c|}{In-domain Testing Set}                             & \multicolumn{4}{c}{(Unseen) Testing Set 1}                                \\ \cline{2-9}
                         & \multicolumn{2}{c}{Left Lung} & \multicolumn{2}{c|}{Right Lung} & \multicolumn{2}{c}{Left Lung} & \multicolumn{2}{c}{Right Lung} \\ \cline{2-9}
                         & DSC (\%)              & NSD (\%)           & DSC (\%)            & NSD (\%)            & DSC (\%)             & NSD (\%)           & DSC (\%)               & NSD (\%)           \\ \hline
StructSeg Lung           &  96.4 $\pm$ \ 1.4 & 74.6 $\pm$ \ 9.1 & 97.3 $\pm$ \ 0.3  & 74.3 $\pm$ \ 7.2  & \textbf{92.2 $\pm$ 19.7} & \textbf{82.0 $\pm$ 15.7} & \textbf{95.5 $\pm$ \ 7.2}  & \textbf{84.2 $\pm$ 11.6} \\
NSCLC Lung               &  95.3 $\pm$ \ 4.9 & 80.2 $\pm$ \ 8.3 & 95.4 $\pm$ 10.9 & 80.7 $\pm$ 10.7 & 57.5 $\pm$ 21.5 & 46.9 $\pm$ 16.9 & 72.2 $\pm$ 15.3 & 51.7 $\pm$ 16.8  \\ \hline
\end{tabular}
\end{table*}

\begin{table}[!htbp]
\caption{Quantitative results (mean $\pm$ standard deviation) of Infection segmentation in Task2.}\label{tab:t2-Infection}
\centering
\setlength{\tabcolsep}{1mm}{
\begin{tabular}{c|cc|cc}
\hline
\multirow{2}{*}{Subtask} & \multicolumn{2}{c|}{In-domain Testing Set} & \multicolumn{2}{c}{(Unseen) Testing Set 1} \\ \cline{2-5}
                         & DSC (\%)              & NSD (\%)              & DSC (\%)               & NSD (\%)            \\ \hline
MSD Lung Tumor           & 67.2 $\pm$ 27.1  &  77.1 $\pm$ 31.4  &  \textbf{25.2 $\pm$ 27.4} & \textbf{26.0 $\pm$ 28.5}   \\
StructSeg Tumor          & 71.3 $\pm$ 29.6  &  70.3 $\pm$ 29.5  & \ 6.0  $\pm$ 12.7 & \ 5.5  $\pm$ 10.7   \\
NSCLC-PE                 & 64.4 $\pm$ 45.5  &  73.7 $\pm$ 12.9  & \ 0.4  $\pm$ \ 0.8  & \ 3.7  $\pm$ \ 4.8    \\ \hline
\end{tabular}}
\end{table}


\begin{table*}[!htbp]
\caption{Quantitative Results of 5-fold cross validation of left lung and right lung segmentation in Task 3.}\label{tab:task3-lung}
\centering
\begin{tabular}{c|c|cc|cc|cc|cc}
\hline
\multicolumn{2}{c|}{\multirow{3}{*}{Subtask}} & \multicolumn{4}{c|}{Validation Set}                              & \multicolumn{4}{c}{Testing Set 1}                                  \\
\cline{3-10}
\multicolumn{2}{c|}{}                         & \multicolumn{2}{c|}{Left Lung} & \multicolumn{2}{c|}{Right Lung} & \multicolumn{2}{c|}{Left Lung} & \multicolumn{2}{c}{Right Lung}  \\
\cline{3-10}
\multicolumn{2}{c|}{}                         & DSC (\%) & NSD (\%)                      & \multicolumn{1}{c|}{DSC (\%)} & NSD (\%)  & DSC (\%)              & NSD (\%)                      & DSC (\%) & NSD (\%)                       \\
\hline
\multirow{6}{*}{StructSeg} & Fold-0          & 96.3 $\pm$ \ 0.1  & 79.9 $\pm$ \ 8.5 & 97.2 $\pm$ \ 0.4 & 73.9 $\pm$ \ 7.0 & 97.4 $\pm$ \ 1.9 & 97.6 $\pm$ \ 2.0 & 90.3 $\pm$ \ 5.9 & 90.8 $\pm$ \ 6.1         \\
                          & Fold-1           & 96.3 $\pm$ \ 1.2  & 73.7 $\pm$ \ 8.4 & 97.1 $\pm$ \ 0.4 & 73.4 $\pm$ \ 7.0 & 97.7 $\pm$ \ 1.3 & 91.0 $\pm$ \ 5.3 & 98.0 $\pm$ \ 1.1 & 91.8 $\pm$ \ 4.9                           \\
                          & Fold-2           & 96.4 $\pm$ \ 1.3  & 74.3 $\pm$ \ 8.5 & 97.2 $\pm$ \ 0.3 & 74.0 $\pm$ \ 6.9 & 96.8 $\pm$ \ 3.1 & 89.4 $\pm$ \ 8.8 & 97.6 $\pm$ \ 2.8 & 90.9 $\pm$ \ 7.8                           \\
                          & Fold-3           & 96.3 $\pm$ \ 1.2  & 73.9 $\pm$ \ 8.5 & 97.2 $\pm$ \ 0.3 & 73.9 $\pm$ \ 7.0 & 96.9 $\pm$ \ 2.5 & 90.7 $\pm$ \ 5.6 & 97.3 $\pm$ \ 2.5 & 91.3 $\pm$ \ 6.3                          \\
                          & Fold-4           & 96.3 $\pm$ \ 1.3  & 73.8 $\pm$ \ 8.9 & 97.2 $\pm$ \ 0.4 & 73.8 $\pm$ \ 7.3 & 97.8 $\pm$ \ 1.3 & 91.6 $\pm$ \ 5.3 & 98.0 $\pm$ \ 1.3 & 92.0 $\pm$ \ 5.7                           \\
\cline{2-10}
                          & Avg              & 96.3 $\pm$ \ 1.2  & 73.9 $\pm$ \ 8.2 & 97.2 $\pm$ \ 0.3 & 73.8 $\pm$ \ 6.7 & \textbf{97.3 $\pm$ \ 2.1} & \textbf{90.6 $\pm$ \ 6.2} & \textbf{97.7 $\pm$ \ 2.1} & \textbf{91.4 $\pm$ \ 6.1}                           \\
\hline
\multirow{6}{*}{NSCLC}     & Fold-0          & 95.7 $\pm$ \ 4.6  & 81.2 $\pm$ \ 7.5 & 95.5 $\pm$ 10.9 & 81.0 $\pm$ 10.9 & 92.7 $\pm$ \ 6.3 & 75.4 $\pm$ 14.5 & 93.0 $\pm$ \ 7.0 & 85.3 $\pm$ 16.0                          \\
                          & Fold-1           & 95.4 $\pm$ \ 5.0  & 80.5 $\pm$ \ 8.7 & 95.2 $\pm$ 11.1 & 80.5 $\pm$ 11.4 & 92.2 $\pm$ \ 7.2 & 73.6 $\pm$ 17.5 & 94.3 $\pm$ \ 3.8 & 76.7 $\pm$ 14.3                         \\
                          & Fold-2           & 95.4 $\pm$ \ 4.9  & 79.7 $\pm$ \ 8.2 & 95.2 $\pm$ 10.9 & 80.0 $\pm$ 11.3 & 94.1 $\pm$ \ 4.1 & 77.4 $\pm$ 12.0 & 93.8 $\pm$ \ 5.8 & 75.6 $\pm$ 16.2                           \\
                          & Fold-3           & 95.4 $\pm$ \ 4.9  & 79.8 $\pm$ \ 7.7 & 94.8 $\pm$ 11.2 & 79.5 $\pm$ 11.6 & 93.6 $\pm$ \ 5.1 & 77.9 $\pm$ 11.6 & 93.6 $\pm$ \ 5.9 & 78.2 $\pm$ 13.3                           \\
                          & Fold-4           & 95.7 $\pm$ \ 4.6  & 81.1 $\pm$ \ 7.9 & 95.5 $\pm$ 10.9 & 80.9 $\pm$ 11.0 & 94.8 $\pm$ \ 3.8 & 80.5 $\pm$ 10.5 & 95.1 $\pm$ \ 3.3 & 80.5 $\pm$ 11.1                           \\
\cline{2-10}
                          & Avg              & 95.5 $\pm$ \ 4.8  & 80.5 $\pm$ \ 8.0 & 95.2 $\pm$ 10.9 & 80.4 $\pm$ 11.2  & 93.5 $\pm$ \ 5.4 & 76.9 $\pm$ 13.3 & 94.0 $\pm$ \ 5.3 & 77.2 $\pm$ 14.1                           \\
\hline
\end{tabular}
\end{table*}

U-Net is often used as a baseline model in existing COVID-19 CT segmentation studies. However, reported results vary a lot even in the same dataset, which make it hard to compare different studies. To standardize the U-Net performance, we build our baselines on nnU-Net that is the most powerful U-Net implementation to the best of our knowledge. To make it comparable between different tasks, we manually adjust the patch sizes and network architectures in Task 2 and Task 3 to be the same as Task 1. Figure \ref{fig:unet} shows details of the U-Net architecture.

During pre-processing, we apply Z-score (mean subtraction and division by standard deviation) to normalize the image intensities. During training, we use the standard training scheme of nnU-Net. For example, The sum between cross entropy and Dice loss is used as the loss function. The optimizer is stochastic gradient descent with initial learning rate (0.01) and a large nesterov momentum (0.99) and, `PolyLR' schedule \cite{DeepLabPAMI} is used to reduce the learning rate. We randomly sample image patches with size $192\times192\times64$.
All training procedures run for a fixed length of 1000 epochs, where each epoch is defined as 250 training iterations (batch size 2). During testing, we use the same patch size with sliding window to infer testing cases and the sliding stride is half the patch size\footnote{We do not apply post-processing to refine the network predictions.}.

\section{Results and Discussion}
This section presents the quantitative segmentation results in each task. For clarity, the first three subsections show the segmentation results on the validation set and the first testing set for each task. We summarize all the segmentation results on the second testing set (MosMed dataset) and compare the results of three tasks in the last subsection.

\subsection{Results of Task 1: Learning with limited annotations}

Table \ref{tab:exp-1} presents average DSC and NSD results of lung and infection of each subtask in Task 1. It can be found that
\begin{itemize}
    \item the average DSC and NSD values among different folds vary greatly. This is because the testing cases in each fold have different degrees of difficulty, which demonstrates that reporting 5-fold cross validation results is necessary to obtain an objective evaluation as 1-fold results may be biased.
    \item promising results for left and right lung segmentation in COVID-19 CT scans can be achieved with as few as four training cases. Models trained for segmenting lung obtain significantly better results compared with those trained for segmenting lung and infection simultaneously.
    \item there is still large room for improving infection segmentation with limited annotations.
\end{itemize}

\begin{figure}[!h]
\begin{center}
   \includegraphics[scale=0.3]{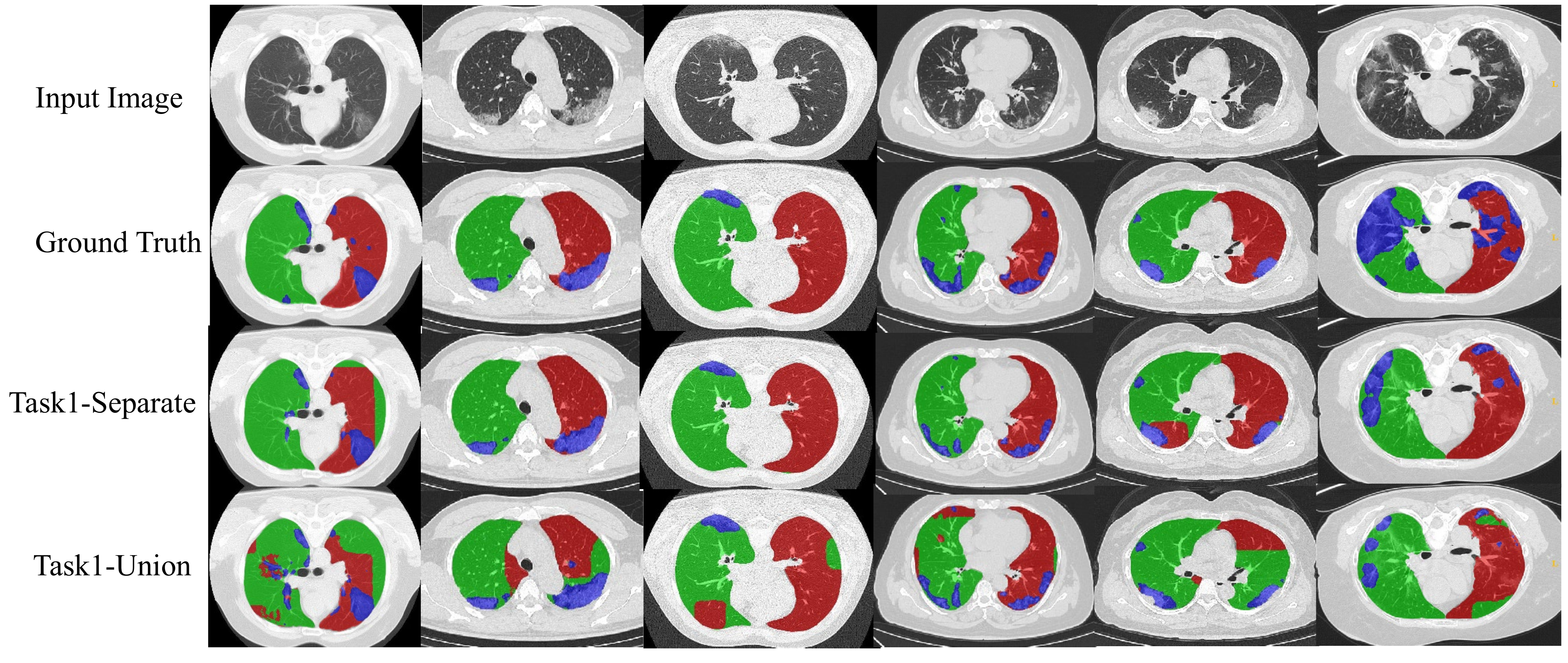}
    \caption{Visualized examples of segmentation results in Task 1. Task1-Separate means the results of training separate networks for lung and infection segmentation. Task1-Union means training a single model for both lung and infection segmentation. The red, green, and blue color denote the left lung, the right lung, and the infection, respectively.}\label{fig:Task1}
\end{center}
\end{figure}

Figure~\ref{fig:Task1} presents some visualized segmentation results in Task 1. It can be found that the separate training manner yields better results, especially for the left lung and right lung segmentation. The union training manner could confuse left and right lung, adversely affecting the infection segmentation. This is because multi-task segmentation is much harder than single task, especially when each separate task is challenging.

\subsection{Results of Task 2: Learning to segment COVID-19 CT scans from non-COVID-19 CT scans}
This task is quite challenging as the model does not see any cases from target domain during training. In other words, the trained models are expected to generalize to the unseen domain (COVID-19 CT).
Table \ref{tab:t2-lung} shows left lung and right lung segmentation results in terms of average and standard deviation values of DSC and NSD. It can be found that
\begin{itemize}
    \item 3D U-Net achieves excellent performance in terms of DSC on the in-domain set. Average NSD values are lower than DSC values, implying that most of the errors come from the boundaries.
    \item the performance on the testing set drops significantly on both subtasks. The performance of the model trained on NSCLC Lung dataset is worse than the model trained on StuctSeg lung. The potential reason could be the difference in the distribution lung appearance is smaller between StructSeg and COVID-19-CT.
\end{itemize}

\begin{table}[!htbp]
\caption{Quantitative Results of 5-fold cross validation of infection segmentation in Task 3.}\label{tab:task3-infection}
\centering
\begin{tabular}{c|c|cc|cc}
\hline
\multicolumn{2}{c|}{\multirow{2}{*}{Subtask}}                                        & \multicolumn{2}{c|}{Validation Set} & \multicolumn{2}{c}{Testing Set 1} \\ \cline{3-6}
\multicolumn{2}{c|}{}                                                                & DSC (\%)              & NSD (\%)              & DSC (\%)            & NSD (\%)            \\ \hline
\multirow{6}{*}{\begin{tabular}[c]{@{}c@{}}MSD\\ Lung\\ Tumor\end{tabular}} & Fold-0 & 67.2 $\pm$ 26.7  & 78.1 $\pm$ 30.8  & 68.0 $\pm$ 22.5& 66.6 $\pm$ 23.7               \\
                                                                            & Fold-1 & 66.3 $\pm$ 26.1  & 76.9 $\pm$ 29.6  & 67.0 $\pm$ 22.0& 65.1 $\pm$ 25.9               \\
                                                                            & Fold-2 & 67.1 $\pm$ 25.4  & 77.4 $\pm$ 27.8  & 63.0 $\pm$ 27.9& 64.4 $\pm$ 28.7               \\
                                                                            & Fold-3 & 63.9 $\pm$ 26.6  & 73.8 $\pm$ 31.1  & 61.7 $\pm$ 24.5& 59.8 $\pm$ 28.5               \\
                                                                            & Fold-4 & 68.0 $\pm$ 25.9  & 78.8 $\pm$ 29.9  & 51.9 $\pm$ 30.6& 50.6 $\pm$ 30.9               \\ \cline{2-6}
                                                                            & Avg    & 66.5 $\pm$ 25.3  & 77.0 $\pm$ 28.9  & 62.3 $\pm$ 25.7& 61.3 $\pm$ 27.6               \\ \hline
\multirow{6}{*}{\begin{tabular}[c]{@{}c@{}}StructSeg\\ Tumor\end{tabular}}  & Fold-0 & 78.2 $\pm$ 14.1  & 75.4 $\pm$ 17.5  & 69.3 $\pm$ 20.5& 68.0 $\pm$ 21.8               \\
                                                                            & Fold-1 & 78.6 $\pm$ 14.0  & 76.1 $\pm$ 18.1  & 68.3 $\pm$ 22.4& 64.8 $\pm$ 26.3               \\
                                                                            & Fold-2 & 77.0 $\pm$ 13.7  & 73.0 $\pm$ 17.0  & 63.6 $\pm$ 25.4& 66.1 $\pm$ 25.5               \\
                                                                            & Fold-3 & 78.8 $\pm$ 13.6  & 76.0 $\pm$ 17.5  & 67.0 $\pm$ 24.1& 66.4 $\pm$ 25.2               \\
                                                                            & Fold-4 & 77.5 $\pm$ 13.7  & 74.6 $\pm$ 18.4  & 52.6 $\pm$ 28.7& 51.2 $\pm$ 28.5               \\ \cline{2-6}
                                                                            & Avg    & 78.0 $\pm$ 13.3  & 75.0 $\pm$ 17.0  & \textbf{64.2 $\pm$ 24.5}& \textbf{63.3 $\pm$ 25.7 }              \\ \hline
\multirow{6}{*}{NSCLC}                                                      & Fold-0 & 65.5 $\pm$ 15.4  & 74.3 $\pm$ 13.2  & 69.2 $\pm$ 20.7& 66.5 $\pm$ 22.4               \\
                                                                            & Fold-1 & 64.7 $\pm$ 15.4  & 73.8 $\pm$ 14.3  & 59.7 $\pm$ 22.7& 55.8 $\pm$ 25.5               \\
                                                                            & Fold-2 & 65.5 $\pm$ 15.2  & 74.8 $\pm$ 13.7  & 61.6 $\pm$ 28.2& 61.7 $\pm$ 29.1               \\
                                                                            & Fold-3 & 64.7 $\pm$ 16.0  & 74.0 $\pm$ 13.6  & 62.7 $\pm$ 25.7& 62.0 $\pm$ 27.7               \\
                                                                            & Fold-4 & 65.2 $\pm$ 15.6  & 74.7 $\pm$ 13.0  & 47.7 $\pm$ 27.2& 46.5 $\pm$ 27.0               \\ \cline{2-6}
                                                                            & Avg    & 65.1 $\pm$ 15.2  & 74.3 $\pm$ 13.2  & 60.2 $\pm$ 25.4& 58.5 $\pm$ 26.7               \\ \hline
\end{tabular}
\end{table}

Table \ref{tab:t2-Infection} shows quantitative infection segmentation results in terms of average and standard deviation values of DSC and NSD. It can be found that
\begin{itemize}
    \item on the in-domain testing set, the performance of lesion segmentation is not as good as the performance of lung segmentation (Table \ref{tab:t2-lung}), which means that tumor segmentation remains a challenging problem. This observation is in line with recent results in MICCAI tumor segmentation challenge, i.e. liver tumor segmentation \cite{bilic2019lits} and kidney tumor segmentation \cite{heller2019kits}.
    \item the models almost fail to predict COVID-19 infections on testing set, which highlights that the lesion appearances differ significantly among lung cancer, pleural effusion, and COVID-19 infections in CT scans.
\end{itemize}

\subsection{Results of Task 3: Learning with both COVID-19 and non-COVID-19 CT scans}
In Task 3, heterogeneous cases from both COVID-19 and non-COVID-19 datasets are leveraged to train for segmenting lung and infections on COVID-19 CT scans. Due to the gap between multiple domains, this data fusion is expected to explore how fusing different annotations influences the model's performance on each individual dataset separately.


Table \ref{tab:task3-lung} presents quantitative 5-fold cross validation results of left lung and right lung segmentation in terms of average DSC and NSD. It can be found that
\begin{itemize}
    \item on the validation set, the average DSC and NSD values are basically consistent with the results in Task 2, achieving high performance with up to 96.4\% in DSC for left lung segmentation and 97.2\% in DSC for right lung segmentation.
    \item on the testing set, however, the average DSC and NSD values slightly drop, indicating that though segmenting the same organ lung on CT scans, there still exists some domain gaps between non-COVID-19 and COVID-19 CT datasets.
\end{itemize}


Table \ref{tab:task3-infection}  present quantitative 5-fold cross validation results of the infection segmentation, it can be found that
\begin{itemize}
    \item even large amounts of lung lesion annotations from non-COVID-19 dataset are used, the variance among the results of 5-fold cross validation is obvious. Thus, reporting 5-fold cross validation results is necessary in this task for reliable and robust evaluation.
    \item compared with the results in Task 2 (Table \ref{tab:t2-Infection}), including four COVID-19 cases bring remarkable improvements with up to 7.5\% in DSC for StructSeg tumor segmentation, and 1.1\% in DSC for NSCLC pleural effusion segmentation, while the performance drops up to 3.3\% in DSC for MSD lung tumor segmentation. These results imply that including few out-of-domain cases in a training set can lead to significant change to the model's performance.
    \item on the testing set, the relative performance drops about 4\%-14\% in DSC and 11\%-16\% in NSD, indicating that simply fusing both COVID-19 and non-COVID-19 cases is still inefficient to segment infections on COVID-19 CT scans. Therefore, there remains much scope for improvement to bridge the gap among many non-COVID-19 lung lesion cases and limited COVID-19 cases by advanced knowledge transfer techniques.
\end{itemize}

\begin{table*}[!htbp]
\caption{Quantitative comparison of COVID-19 CT lung and infection segmentation results among different tasks on testing set in terms of average DSC and NSD values of all testing cases.}\label{tab:task-PK}
\centering
\begin{tabular}{c|cc|cc|cc|cc}
\hline
\multirow{2}{*}{Subtask} & \multicolumn{2}{c|}{Left Lung}                    & \multicolumn{2}{c|}{Right Lung}                   & \multicolumn{2}{c|}{Infection (COVID-19-CT-Seg)}   & \multicolumn{2}{c}{Infection (MosMed)}          \\
\cline{2-9}
                         & DSC (\%)                     & NSD (\%)                     & DSC (\%)                     & NSD (\%)                     & DSC (\%)                       & NSD (\%)                       & \multicolumn{1}{c}{DSC (\%)} & \multicolumn{1}{c}{NSD (\%)}  \\
\hline
Task1-Separate           & 85.8 $\pm$ 10.5         & 71.2 $\pm$ 13.8         & 87.9 $\pm$ 9.3           & 74.8 $\pm$ 11.9         & \textbf{67.3 $\pm$ 22.3}  & \textbf{70.0 $\pm$ 24.4}  & \textbf{58.8 $\pm$ 20.6}   & \textbf{66.4 $\pm$ 20.3}   \\
Task1-Union              & 64.6 $\pm$ 26.4         & 51.1 $\pm$ 23.4         & 75.0 $\pm$ 16.8         & 57.7 $\pm$ 17.4         & 61.0 $\pm$ 26.2           & 61.8 $\pm$ 27.4           & 48.2 $\pm$ 22.1          & 41.4 $\pm$ 19.1 \\
\hline
Task2-MSD                & \multicolumn{2}{c|}{-}                            & \multicolumn{2}{c|}{-}                            & 25.2 $\pm$ 27.4           & 26.0 $\pm$ 28.5           & 16.2 $\pm$ 23.2                        & 17.5 $\pm$ 23.4                         \\
Task2-StructSeg          & 92.2 $\pm$ 19.7         & 82.0 $\pm$ 15.7         & 95.5 $\pm$ 7.2           & 84.2 $\pm$ 11.6         &  6.0 $\pm$ 12.7            &  5.5 $\pm$ 10.7            & 2.6 $\pm$ 9.5                        & 3.3 $\pm$ 9.9                         \\
Task2-NSCLC              & 57.5 $\pm$ 21.5         & 46.9 $\pm$ 17.0         & 72.2 $\pm$ 15.3         & 51.7 $\pm$ 16.8         &  0.4 $\pm$ 0.9              &  3.7 $\pm$ 4.8              & 0.0 $\pm$ 0.0                        & 0.5 $\pm$ 1.4                         \\
\hline
Task3-MSD                & \multicolumn{2}{c|}{-}                            & \multicolumn{2}{c|}{-}                            & 62.3 $\pm$ 25.7           & 61.3 $\pm$ 27.6           & 39.2 $\pm$ 30.6                        & 41.3 $\pm$ 30.5                         \\
Task3-StructSeg          & \textbf{97.3 $\pm$ 2.1}  & \textbf{90.6 $\pm$ 6.2}  & \textbf{97.7 $\pm$ 2.1}  & \textbf{91.4 $\pm$ 6.1}  & 64.2 $\pm$ 24.5           & 63.3 $\pm$ 25.7           & 44.3 $\pm$ 25.3                        & 49.1 $\pm$ 25.8                         \\
Task3-NSCLC              & 93.5 $\pm$ 5.4           & 76.9 $\pm$ 13.3         & 94.0 $\pm$ 5.3           & 77.2 $\pm$ 14.1         & 60.2 $\pm$ 25.4           & 58.5 $\pm$ 26.7           & 30.1 $\pm$ 26.7                        & 33.4 $\pm$ 27.1                         \\
\hline
\end{tabular}
\end{table*}

\begin{table*}[!htbp]
\caption{Quantitative comparison of COVID-19 CT lung and infection segmentation results among different tasks on testing set in terms of average sensitivity and specificity values of all testing cases.}\label{tab:sen-spe-PK}
\centering
\setlength{\tabcolsep}{1mm}{
\begin{tabular}{c|cc|cc|cc|cc}
\hline
\multirow{2}{*}{Subtask} & \multicolumn{2}{c|}{Left Lung}                    & \multicolumn{2}{c|}{Right Lung}                   & \multicolumn{2}{c|}{Infection (COVID-19-CT-Seg)}   & \multicolumn{2}{c}{Infection (MosMed)}          \\
\cline{2-9}
                         & Sensitivity (\%)                     & Specificity (\%)                     & Sensitivity (\%)                     & Specificity (\%)                     & Sensitivity (\%)                       & Specificity (\%)                       & \multicolumn{1}{c}{Sensitivity (\%)} & \multicolumn{1}{c}{Specificity (\%)}  \\
\hline
Task1-Separate           & 86.2 $\pm$ 11.6         & 99.2 $\pm$ 1.5         & 89.7 $\pm$ 12.3           & 99.1 $\pm$ 7.3         & 62.0 $\pm$ 23.7  & 99.9 $\pm$ 15.9  & 57.5 $\pm$ 23.8   & 99.9 $\pm$ 0.0  \\
Task1-Union              & 67.0 $\pm$ 28.5         & 98.8 $\pm$ 1.2         & 81.4 $\pm$ 19.5        & 98.2 $\pm$ 1.6         & 62.8 $\pm$ 27.1           & 99.7 $\pm$ 3.0           & 60.1 $\pm$ 24.3          & 99.9 $\pm$ 0.2 \\
\hline
Task2-MSD                & \multicolumn{2}{c|}{-}                            & \multicolumn{2}{c|}{-}                            & 18.6 $\pm$ 23.1           & 100 $\pm$ 0.1           & 13.1 $\pm$ 22.9                        & 100 $\pm$ 0.0                         \\
Task2-StructSeg          & 91.7 $\pm$ 20.8         & 99.9 $\pm$ 0.1         & 95.3 $\pm$ 10.2           & 99.8 $\pm$ 0.2         &  1.2 $\pm$ 2.4            &  100 $\pm$ 0.0            & 1.8 $\pm$ 6.9                        & 100 $\pm$ 0.0                         \\
Task2-NSCLC             & 47.6 $\pm$ 23.4         & 99.4 $\pm$ 0.6         & 81.6 $\pm$ 21.7           & 97.4 $\pm$ 1.9         &  37.6 $\pm$ 26.6              &  100 $\pm$ 0.0              & 0.0 $\pm$ 0.0                        & 100 $\pm$ 0.0                         \\
\hline
Task3-MSD                & \multicolumn{2}{c|}{-}                            & \multicolumn{2}{c|}{-}                            & 63.0 $\pm$ 27.4           & 99.8 $\pm$ 0.3           & 36.4 $\pm$ 32.7                        & 100 $\pm$ 0.0                         \\
Task3-StructSeg          & 97.5 $\pm$ 2.7  & 99.9 $\pm$ 0.2  & 98.0 $\pm$ 2.0  & 99.8 $\pm$ 0.2  & 64.8 $\pm$ 25.3           & 99.8 $\pm$ 0.3           & 42.2 $\pm$ 29.5                        & 100 $\pm$ 0.1                         \\
Task3-NSCLC              & 93.4 $\pm$ 7.1           & 99.7 $\pm$ 0.3         & 96.1 $\pm$ 3.4           & 99.5 $\pm$ 0.6         & 62.4 $\pm$ 26.7           & 99.7 $\pm$ 0.4           & 24.9 $\pm$ 25.9                        & 100 $\pm$ 0.0                         \\
\hline
\end{tabular}}
\end{table*}

\begin{table*}[!htbp]
\centering
\setlength{\tabcolsep}{1mm}{
\begin{tabular}{c|cc|cc|cc|cc}
\hline
\multirow{2}{*}{Subtask} & \multicolumn{2}{c|}{Left Lung}                    & \multicolumn{2}{c|}{Right Lung}                   & \multicolumn{2}{c|}{Infection (COVID-19-CT-Seg)}   & \multicolumn{2}{c}{Infection (MosMed)}          \\
\cline{2-9}
                         & Recall (\%)                     & Precision (\%)                     & Recall (\%)                     & Precision (\%)                    & Recall (\%)                       & Precision (\%)                       & \multicolumn{1}{c}{Recall (\%)} & \multicolumn{1}{c}{Precision (\%)}  \\
\hline
Task1-Separate           & 86.9 $\pm$ 11.2         & 79.6 $\pm$ 19.6         & 90.2 $\pm$ 13.4           & 77.6 $\pm$ 14.7         & 62.0 $\pm$ 23.7  & 84.0 $\pm$ 19.6  & 57.5 $\pm$ 23.8   & 67.9 $\pm$ 21.7  \\
Task1-Union              & 67.0 $\pm$ 28.5         & 71.6 $\pm$ 23.9         & 81.4 $\pm$ 19.5        & 72.8 $\pm$ 17.9         & 62.8 $\pm$ 27.1           & 74.1 $\pm$ 26.2           & 60.1 $\pm$ 24.3          & 57.7 $\pm$ 28.2 \\
\hline
Task2-MSD                & \multicolumn{2}{c|}{-}                            & \multicolumn{2}{c|}{-}                            & 18.6 $\pm$ 23.1           & 78.7 $\pm$ 34.3           & 13.1 $\pm$ 22.9                        & 47.9 $\pm$ 38.6                         \\
Task2-StructSeg          & 91.7 $\pm$ 20.8         & 97.0 $\pm$ 2.1         & 95.3 $\pm$ 10.2           & 96.3 $\pm$ 3.2         &  1.2 $\pm$ 2.4            &  22.6 $\pm$ 32.8            & 1.8 $\pm$ 6.9                        & 9.3 $\pm$ 22.3                         \\
Task2-NSCLC              &  47.6 $\pm$ 23.4              &  80.2 $\pm$ 15.0                            & 81.6 $\pm$ 21.7                        & 67.3 $\pm$ 11.7        &  20.7 $\pm$ 49.8              &  6.8 $\pm$ 11.3             &  9.0 $\pm$ 3.5              &  17.6 $\pm$ 71.8      \\
\hline
Task3-MSD                & \multicolumn{2}{c|}{-}                            & \multicolumn{2}{c|}{-}                            & 63.0 $\pm$ 27.4           & 75.0 $\pm$ 26.1           & 36.4 $\pm$ 32.7                        & 61.4 $\pm$ 30.4                         \\
Task3-StructSeg          & 97.5 $\pm$ 2.7  & 97.2 $\pm$ 2.8  & 98.0 $\pm$ 2.0  & 97.4 $\pm$ 2.7  & 64.8 $\pm$ 25.3           & 76.8 $\pm$ 24.8           & 42.2 $\pm$ 29.5                        & 60.7 $\pm$ 26.0                         \\
Task3-NSCLC              & 93.7 $\pm$ 6.8           & 93.6 $\pm$ 5.9         & 95.9 $\pm$ 3.8           & 92.3 $\pm$ 7.7         & 62.4 $\pm$ 27.3           & 73.1 $\pm$ 26.9           & 24.9 $\pm$ 25.9                        & 61.4 $\pm$ 33.0                         \\
\hline
\end{tabular}}
\end{table*}

\subsection{Comparison among different tasks}

\begin{figure}[htbp]
	\begin{center}
    	\subfloat{\includegraphics[scale=0.23]{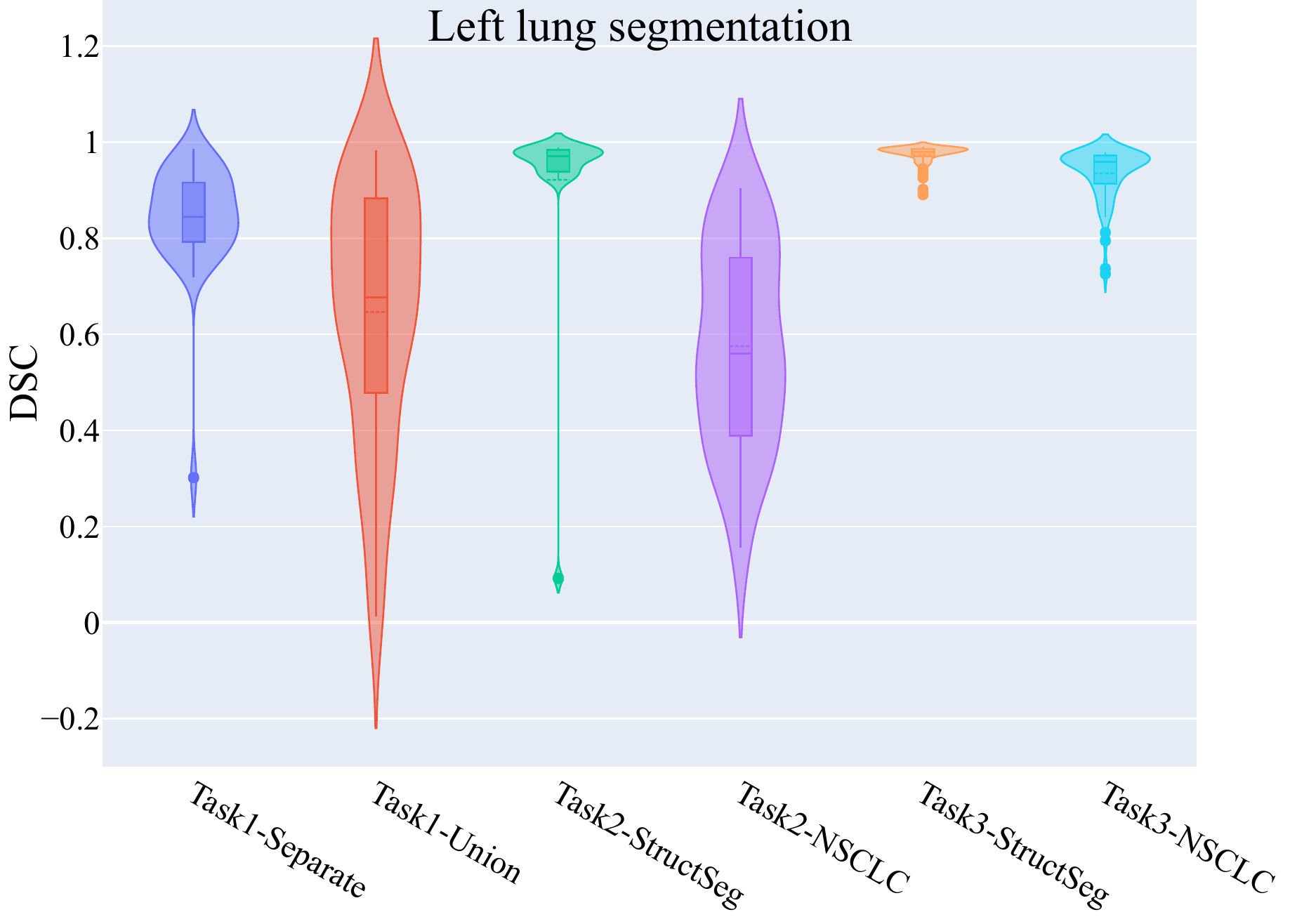}}
    	\subfloat{\includegraphics[scale=0.23]{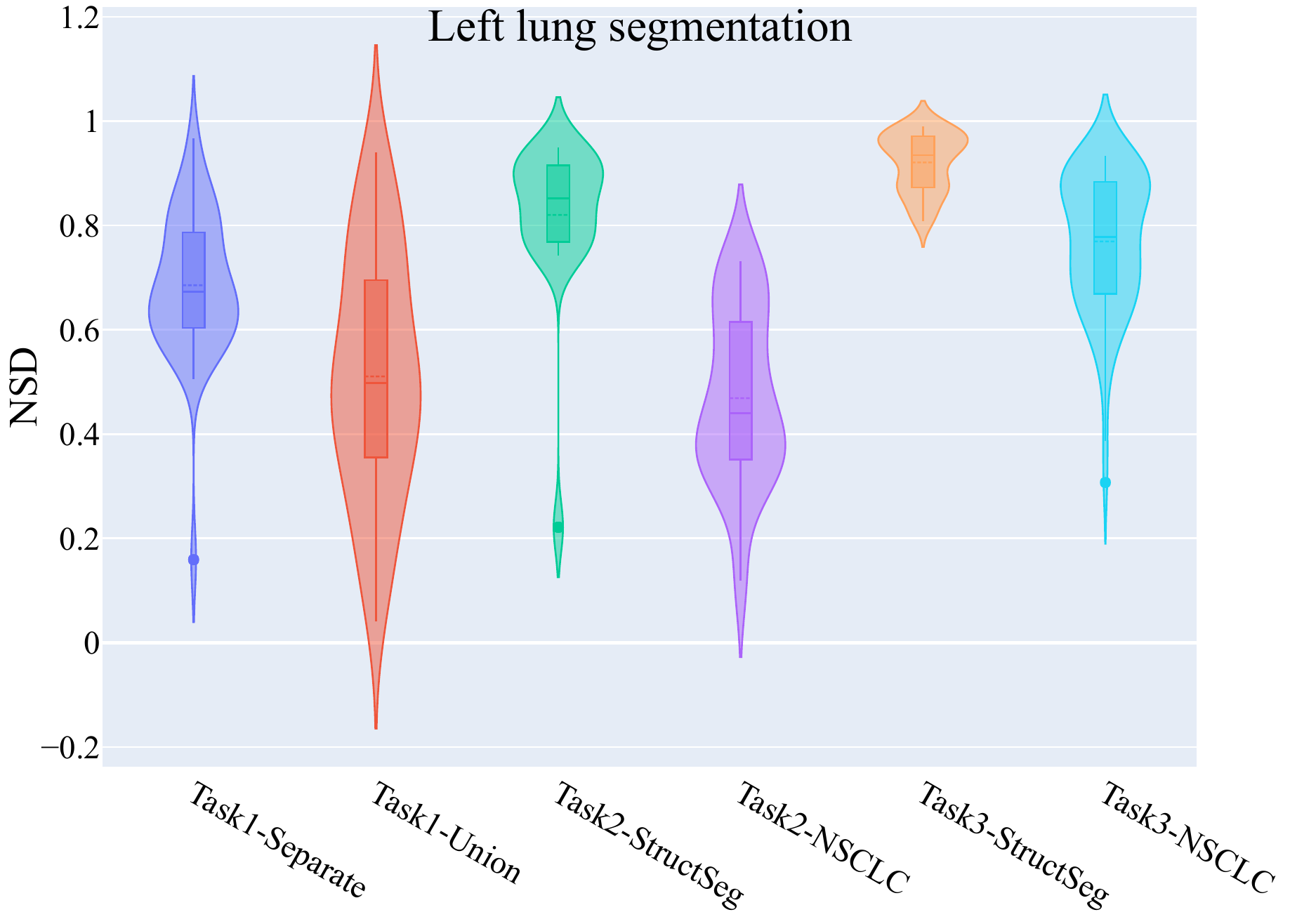}}
    	\subfloat{\includegraphics[scale=0.23]{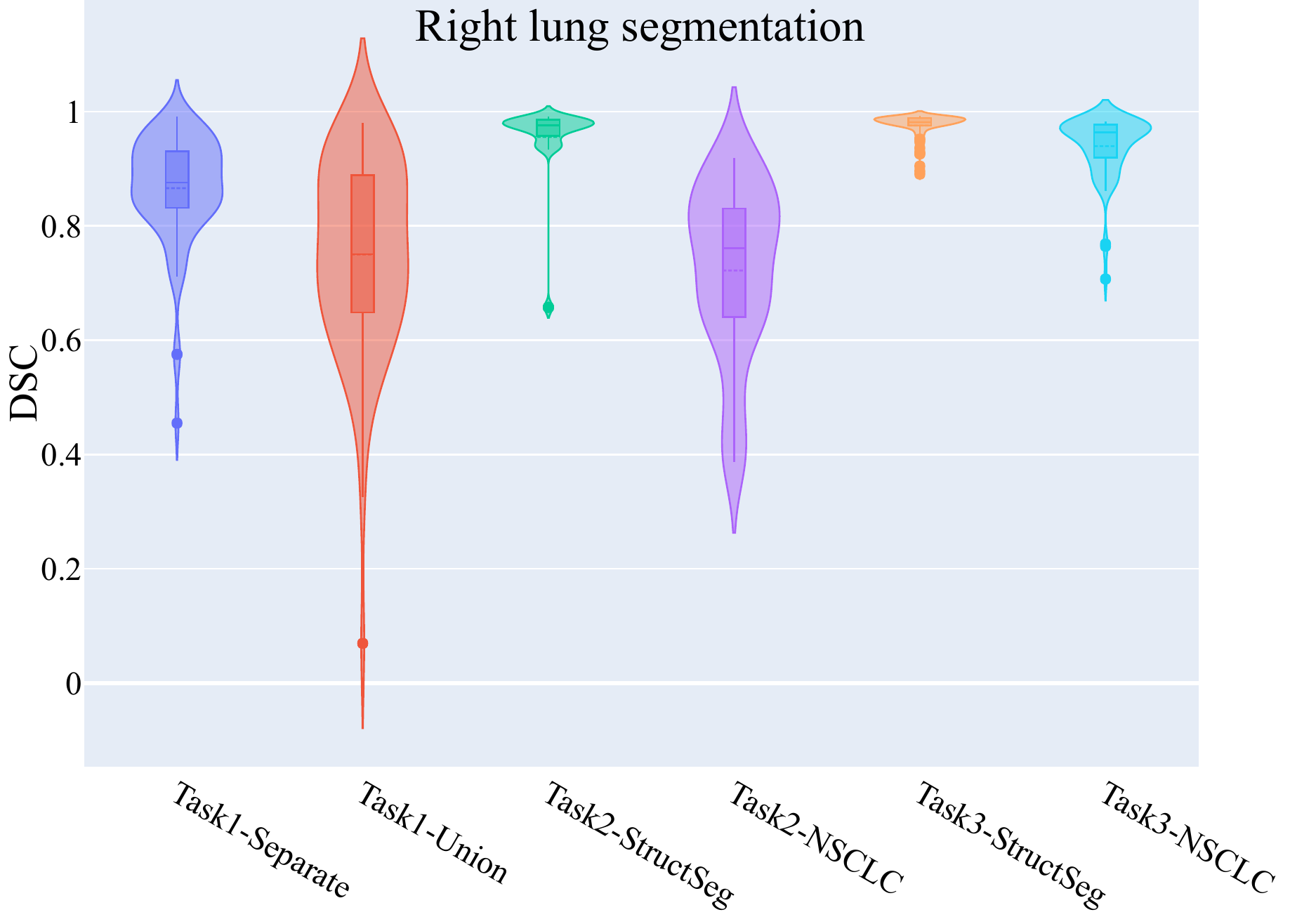}}\\
    	\subfloat{\includegraphics[scale=0.23]{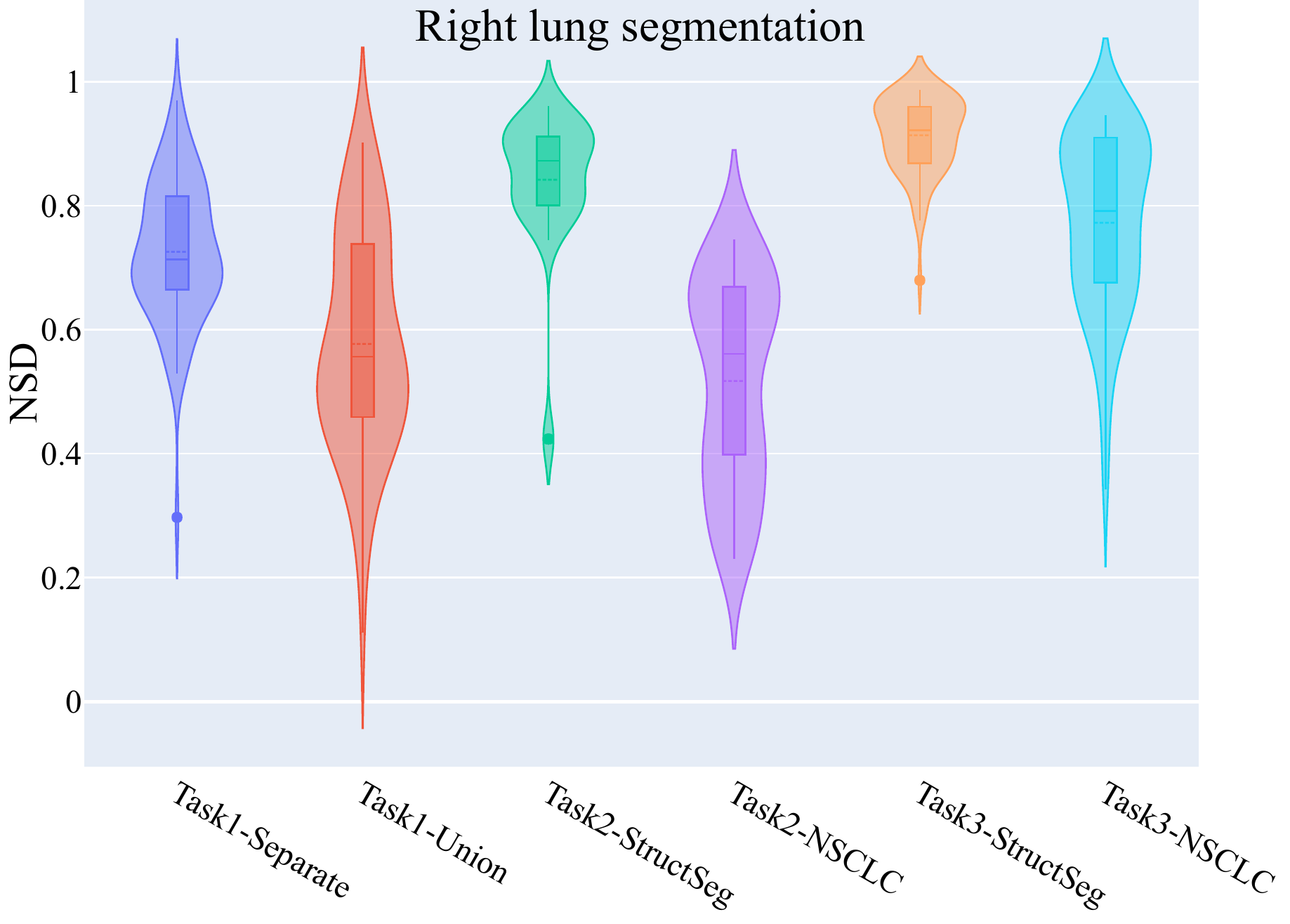}}
    	\subfloat{\includegraphics[scale=0.23]{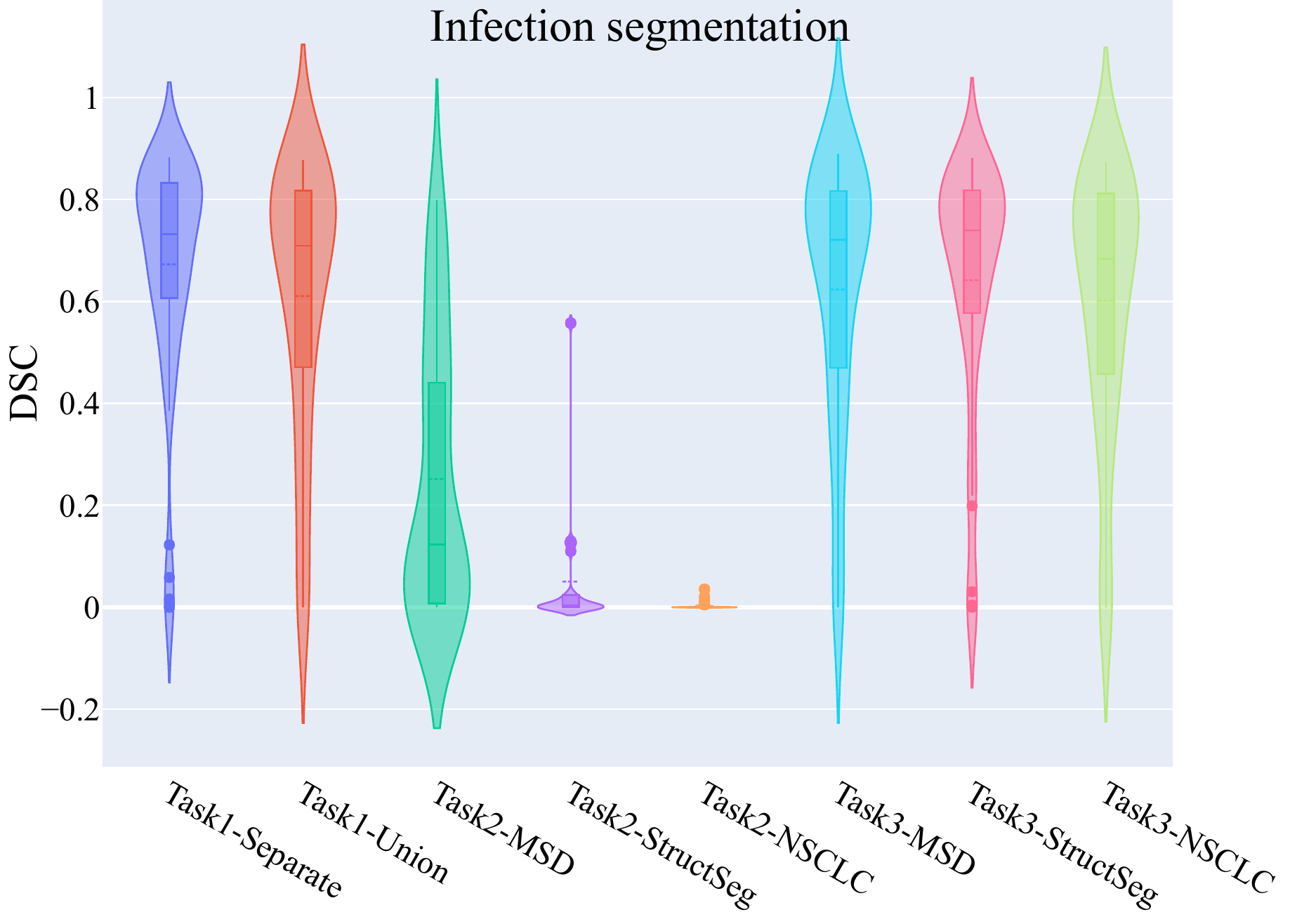}}
    	\subfloat{\includegraphics[scale=0.23]{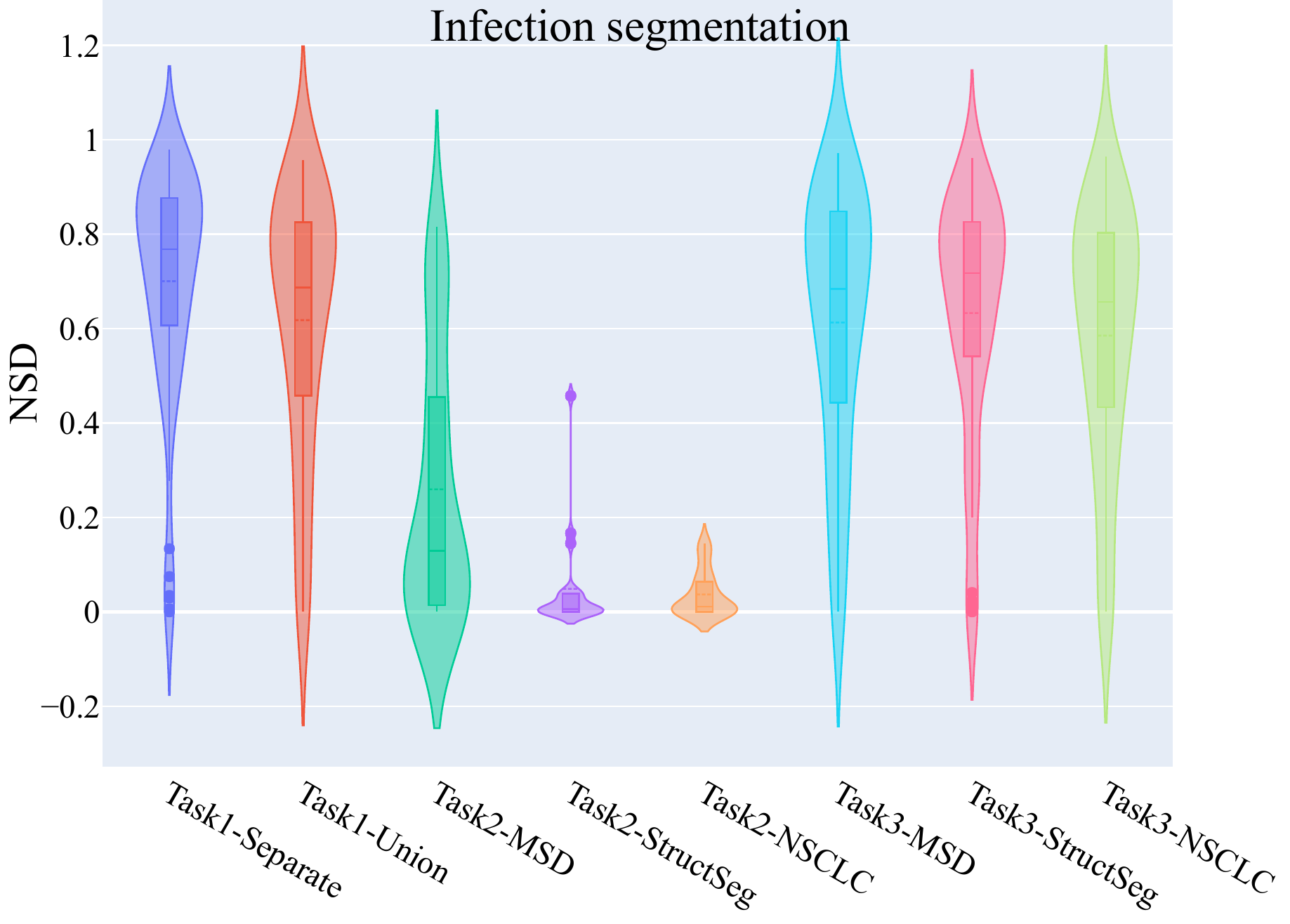}}\\
	\end{center}
	\caption{The violin plots present the performances (DSC or NSD) of different methods for left lung, right lung, and infection segmentation on COVID-19-CT-Seg testing set.}\label{fig:violin}
\end{figure}

Tasks 1-3 correspond to different strategies for lung and infection segmentation in COVID-19 CT scans with limited in-domain training cases and out-of-domain datasets. The testing cases are the same in Tasks 1-3, so it is feasible and reasonable to conduct comparison among different tasks.
Table~\ref{tab:task-PK} presents quantitative results on all testing sets of the three tasks in terms of average DSC and NSD. Task 1-Separate and -Union denote training the network to segment lung and infection separately and simultaneously, respectively. Figure \ref{fig:violin} shows the violin plots of left lung, right lung, and infection segmentation results on COVID-19-CT-Seg dataset in terms of DSC and NSD for all tasks. The violin plot shows not only the summary statistics such as median and interquartile ranges, but also the entire distribution of the quantitative results.

For lung segmentation,
\begin{itemize}
    \item Task 3 achieves the best performance reaching up to 97.3\% in DSC and 90.6\% in NSD for left lung segmentation, and 97.7\% in DSC and 91.4\% in NSD for right lung segmentation.
    \item comparison of the results between Task 1 and Task 2, StructSeg achieves significantly better performance than NSCLC on the unseen testing set, indicating that the domain gap of lung between StructSeg and COVID-19 CT is smaller. The results on NSCLC dataset with $300+$ training cases are worse than the results on COVID-19 dataset with four training cases, indicating that the number of training cases is not the most important while including the in-domain data during learning process is much more important.
    \item comparison of the results between Task 1 and Task 3, adding both out-of-domain datasets (StructSeg and NSCLC) can boost performance of left and right lung segmentation. The results imply that existing non-COVID CT annotations can be used to assist lung segmentation in COVID-19 CT scans. This finding is very encouraging for fast developing a COVID-19 lung segmentation system with limited data especially COVID-19 CT annotations are scarce at present.
    \item comparison of the results between Task 2 and Task 3, adding COVID-19 cases into training set can obtain performance gains on both subtasks (StructSeg and NSCLC), especially for NSCLC that achieves a significant increase of up to 34\% in DSC for left lung segmentation. This result highlights that including few COVID-19 cases in training set is critical for lung segmentation.
\end{itemize}


For infection segmentation,
\begin{itemize}
    \item Task 1 achieves the best performance in both two testing set reaching up to 67.3\% in DSC and 70.0\% in NSD.
    \item comparison of the results between Task 1 and Task 2, using four COVID-19 CT training cases obtains significant better results than using other lung lesion cases (i.e. lung cancer and pleural effusion), which highlights the importance of in-domain data when developing the COVID-19 infection segmentation system.
    \item comparison of the results between Task 1 and Task 3, adding many ($40-62$) non-COVID-19 cases during training degrades instead of increasing the performance, implying that out-of-domain lung lesion data can bias the model's representation ability for COVID-19 infection segmentation.
    \item comparison of the results between Task 2 and Task 3: using only out-of-domain cases can not predict COVID-19 infections while adding a few COVID-19 cases can significantly boost the performance, which highlights that including a few in-domain cases during training is very critical for developing infection segmentation models.
\end{itemize}

\begin{figure*}[!htbp]
\begin{center}
   \includegraphics[scale=0.18]{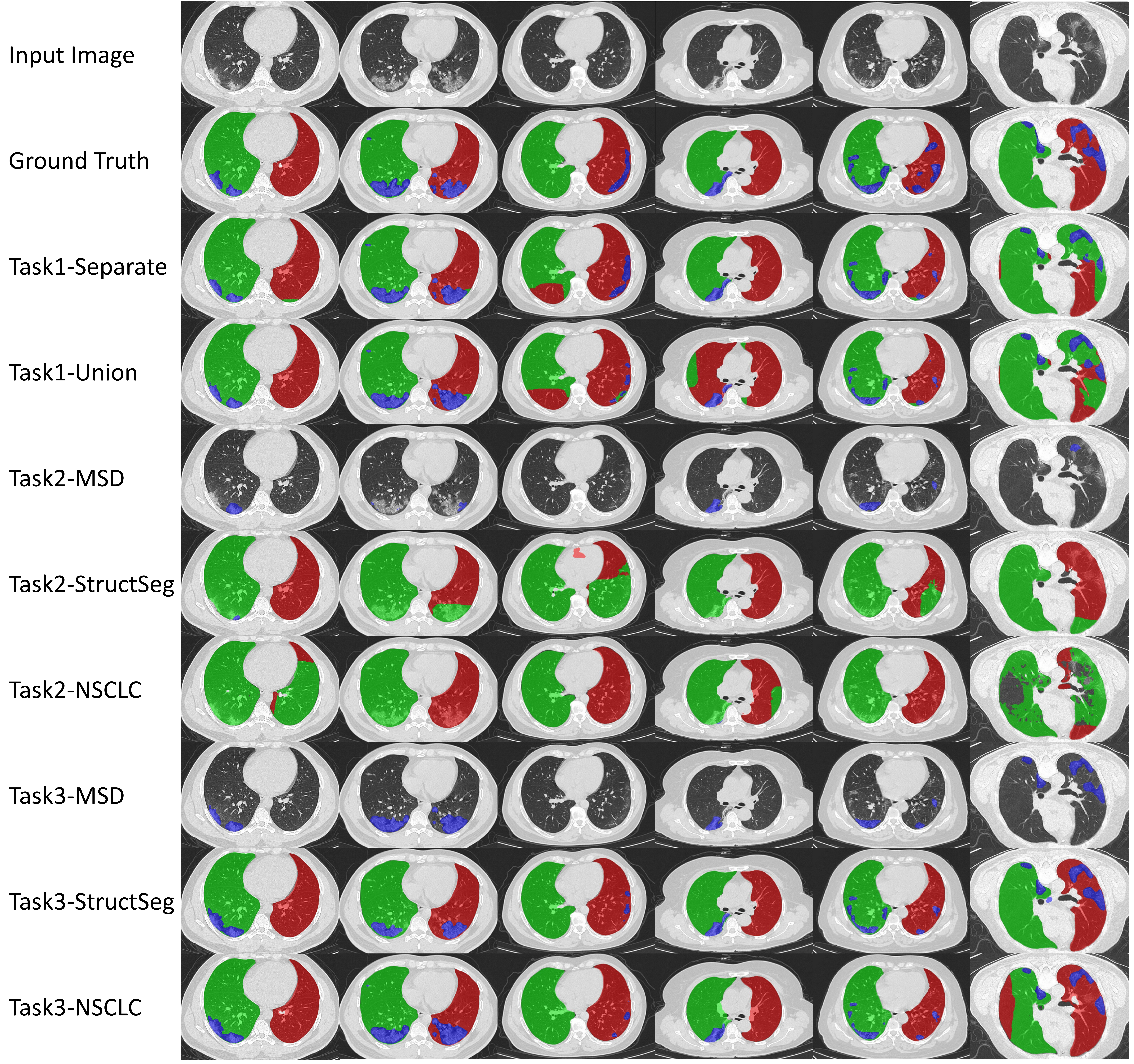}
    \caption{Visualized examples of segmentation results in different tasks. The red, green, and blue color denote the left lung, the right lung, and the infection, respectively.}\label{fig:seg-eg}
\end{center}
\end{figure*}

In addition, it can be found that the infection segmentation performance is lower than lung segmentation in all the experiments. There are two main reasons. First, the task setting is few-shot learning, where only limited labelled cases are allowed to train networks.
Second, compared with the lung, the infection areas are relatively small and most of the infections have weak boundaries. Thus, infection segmentation is much more challenging than lung segmentation.
Table~\ref{tab:sen-spe-PK} shows the recall and precision of the segmentation results in different tasks. It can be found that the infection segmentation results have relatively high precision scores and low recall scores, indicating that the model fails more in detecting all infections. Lung segmentation results achieve better recall and precision scores because the lung tissues have more clear boundaries and larger sizes.
Figure~\ref{fig:seg-eg} presents some visualized segmentation results in different tasks. We can find that the lung segmentation mostly failed due to the confusion between the left lung and the right lung because they share very similar appearances. The large and obvious infections have better segmentation results. However, the infections in small sizes or with weak boundaries are the most often failed cases.

In summary, using the non-COVID-19 chest CT dataset can directly improve the lung segmentation results significantly, but it has few positive impacts on the infection segmentation because of the large domain gap. We found that the infections with good contrast and clear boundaries can be well segmented even with only four training cases. However, the trained models often miss the small infections and weak-boundary infections, indicating that it is hard for the models to capture these features during learning process. Our results also highlight the need of efficient learning methods with limited annotated data. Although including more training cases could be a simple and direct way to boost the infection segmentation performance, one should keep in mind that, in clinical practice, it is impractical to manually annotate many 3D COVID-19 CT scans for each medical center especially when the radiologists are busy with fighting the pandemic. This is our main motivation to set up the data-efficient learning benchmark.

There are some potential solutions for the performance improvements. For example, in few-shot learning task (Task 1), one can use more advanced data augmentation methods~\cite{mengzhang2020dataAug} to increase the training set. In addition, designing task-specific data augmentation methods is also a promising solution, such as augment more cases with small and weak boundary infections.
In domain generalization task (Task 2), the models trained with only non-COVID-19 dataset fail to segment infections. One can introduce more advanced model-agnostic learning methods to handle the domain gap, such as meta-learning~\cite{dou2019domain,khandelwal2020domain}.
In knowledge transfer task (Task 3), simply fusing non-COVID-19 and COVID-19 dataset with the SOTA network could bias the model to learn more non-COVID-19 features. One can use more robust and powerful domain adaptation methods to handle heterogeneous datasets, such as self-supervised learning~\cite{model-genesis}, cross-domain adaptation~\cite{chen2020DA, liu2020ms}.

\subsection{Limitation}
One possible limitation is that the number of cases in our dataset is relatively small. However, this paper focuses on how to learn from limited training cases. Thus, we believe the number of training cases is acceptable for our benchmark tasks. More importantly, this benchmark is also applicable to general small-sample learning problems.
In addition, the number of cases (16 or 20 cases) for the first testing set (COVID-19-CT-Seg) and 50 cases for the second testing set (MosMed) is comparable with recent MICCAI 2020 segmentation challenges. For example, StructSeg (Automatic Structure Segmentation for Radiotherapy Planning Challenge 2020) has \emph{10} testing cases \cite{StructSeg2020} and ASOCA (Automated Segmentation Of Coronary Arteries) has \emph{20} testing cases \cite{ASOCA2020}. Another limitation is that the innovative methodology contribution is limited. However, this is not the primary goal in this paper. Rather, our main goal is to lay the foundation for future work in learning with limited annotated data, and we believe the dataset and the tasks mentioned in the benchmark could attract attentions in the field.


\section{Conclusion}
With the outbreak of COVID-19 around the world, it has become an emergency need to develop deep learning-based COVID-19 image analysis tools with limited data.
To promote the research towards this goal, in this paper, we created a COVID-19 CT dataset, established three segmentation benchmark tasks, and provided $40+$ baselines models based on state-of-the-art segmentation architectures.
All the related results are publicly available at \url{https://github.com/JunMa11/COVID-19-CT-Seg-Benchmark}.
The unified task settings can make the comparison between studies more feasible. The public baselines can save model training time for researchers so that they can focus on developing their own methods.
We hope this work could accelerate the COVID-19 studies on learning with limited data for years to come.

\section*{Acknowledgment}
This project is supported by the National Natural Science Foundation of China (No. 91630311, No. 11971229).
We would like to thank all the organizers of MICCAI 2018 Medical Segmentation Decathlon, MICCAI 2019 Automatic Structure Segmentation for Radiotherapy Planning Challenge, the Coronacases Initiative and Radiopaedia, and Moscow municipal hospitals for their publicly available lung CT dataset. We also thank Joseph Paul Cohen for providing the convenient download link of 20 COVID-19 CT scans, and all the contributors of NSCLC and COVID-19-Seg-CT dataset for providing the annotations of lung, pleural effusion and COVID-19 infection.
We are grateful to the contributors of the two great COVID-19 related resources: COVID19 imaging AI paper list\footnote{https://github.com/HzFu/COVID19\_imaging\_AI\_paper\_list} and MedSeg\footnote{http://medicalsegmentation.com/covid19/} for their timely update about COVID-19 publications and datasets.
Last but not least, we thank the anonymous reviewers, Chen Chen, Xin Yang, and Yao Zhang for their important and valuable feedback on this work.







\bibliographystyle{IEEEtran}
\bibliography{COVID19Ref}
%




\end{document}